\documentclass[12pt]{article}
\usepackage{color}
\usepackage{amsmath}
\usepackage{amssymb}
\usepackage{graphicx}
\usepackage{cite}

\makeatletter
\usepackage{amsthm}\usepackage{amsfonts}%\usepackage{appendix}
\usepackage{slashed}\usepackage{afterpage}%\usepackage{cite}
\usepackage{pdfsync}\usepackage{color}\definecolor{verde}{cmyk}{.83,.21,1,.08}

\def\be{\begin{equation}}
\def\ee{\end{equation}}
\def\bea{\begin{eqnarray}}
\def\eea{\end{eqnarray}}

\newcommand{\eqn}[1]{(\ref{#1})}

\makeatother

\begin{document}
\begin{titlepage}

\begin{flushright}
ICCUB-14-062
\par\end{flushright}

\begin{center}
\baselineskip=24pt
\par\end{center}

\begin{center}
\textbf{\Large Unification of Coupling Constants, Dimension six Operators
and the Spectral Action}{\Large {} }
\par\end{center}

\begin{center}
\baselineskip=14pt
\par\end{center}

\begin{center}
\vspace{1cm}

\par\end{center}

\begin{center}
{Agostino Devastato$^{1,2}$, Fedele Lizzi$^{1,2,3}$, Carlos Valc\'arcel Flores$^{4}$
and Dmitri Vassilevich$^{4}$} \\ $^{1}$\textit{Dipartimento
di Fisica, Universit\`a di Napoli }\textsl{Federico II} \\ $^{2}$\textit{INFN,
Sezione di Napoli}\\
 \textit{Monte S.~Angelo, Via Cintia, 80126 Napoli, Italy} \\
$^{3}$ \textit{Departament de Estructura i Constituents de la Mat\`eria,
}\\
 \textit{Institut de Ci\`encies del Cosmos, Universitat de Barcelona,}\\
 \textit{ Barcelona, Catalonia, Spain} \\ $^{4}$ \textit{CMCC-Universidade
Federal do ABC, Santo Andr\`e, S.P., Brazil } \\
 \texttt{\small devastato@na.infn.it, fedele.lizzi@na.infn.it,
valcarcel.flores@gmail.com, dvassil@gmail.com%
}{\small {} }
\par\end{center}

\vskip 2 cm
\begin{abstract}
We investigate whether inclusion of dimension six terms in the Standard Model lagrangean may cause the unification of the coupling constants
at a scale comprised between $10^{14}$ and $10^{17}$ GeV.
Particular choice of the dimension 6 couplings is motivated by the spectral action. Given the theoretical and phenomenological constraints, as well
as recent data on the Higgs mass, we find that the unification is indeed possible, with a lower unification scale slightly favoured.
\end{abstract}
\end{titlepage}

\section{Introduction}

The coupling constants of the three gauge interactions run with energy~\cite{particledata}.
The ones relating to the nonabelian symmetries are relatively strong
at low energy, but decrease, while the abelian interaction increases.
At an energy comprised between $10^{13}-10^{17}$~GeV their values
are very similar, around $0.52$, but, in view of present data, and
\emph{in absence of new physics}, they fail to meet at a single scale.
Here by absence of new physics we mean extra terms in the Lagrangian
of the model. The extra terms may be due for example to the presence of new particles,
or new interaction. A possibility could be supersymmetric models which
can alter the running and cause the presence of the unification point~\cite{Susy}.

The standard model of particle interaction coupled with gravity may be explained to some extent as a particular for of Noncommutative, or spectral geometry, see for example~\cite{WalterBook} for a recent introduction.
The principles of noncommutative geometry are rigid enough
to restrict gauge groups and their fermionic representations, as well as
to produce a lot of relations between bosonic couplings when applied
on (almost) commutative spaces. All these restrictions and relations are
surprisingly well compatible with the Standard Model, except that the
Higgs field comes out too heavy, and that the unification point of gauge couplings is not exactly found. We have nothing new to say about the first problem, which has been solved in \cite{Stephan,coldplay,resilience,CCvS,BoyleFarnsworthsigma} with the introduction of a new scalar field $\sigma$ suitably coupled to the Higgs field, but we shall address the second one.

Some years ago the data were compatible with the presence of a single
unification point $\Lambda$. This was one of the motivations
behind the building of grand unified theories. Such a feature is however
desirable even without the presence of a larger gauge symmetry group
which breaks to the standard model with the usual mechanisms.
In particular, the approach to field theory, based on noncommutative
geometry and spectral physics~\cite{spectralaction}, needs a scale to regularize the theory. In this respect, the finite mode
regularization~\cite{fujikawa,AndrianovBonora1,AndrianovBonora2}
is ideally suited. In this case $\Lambda$ is also the field theory
cutoff. In fact using this regularization it is possible to generate the bosonic action starting form the fermionic one~\cite{AndrianovLizzi,AndrianovKurkovLizzi,KurkovLizzi}, or describe induced gravity on an equal footing with the anomaly-induced effective action~\cite{KurkovSakellariadou}

The aim of this paper is to investigate whether the presence of higher
dimensional terms in the standard model action $-$ dimension six in particular $-$ may cause the unification of the coupling constants.
The paper may be read in two contexts: as an application of the spectral action, or independently on it, from a purely phenomenologically point of view.

From the spectral point of view, the spectral action~\cite{spectralaction} is solved
as a heath kernel expansion in powers in the inverse of an energy scale.
The terms up to dimension four reproduce the standard model qualitatively,
but the theory is valid at a scale in which the couplings are equal.
The expansion gives, however, also higher dimensional terms, suppressed
by the power of the scale, and depending on the details of the cutoff.
This fixes relations among the coefficients of the new terms. The
analysis of this paper gives the conditions under which the spectral
action can predict the unification of the three gauge coupling constants.

On the other side, it is also possible to read the paper at a purely
phenomenological level, using the spectral action as input only for
the choice of the subset of all possible higher dimensions terms in
the action, and as a guide for the setting of the low energy values for the couplings of the coefficients of the extra terms. We show that the presence of these terms enables the possibility
of a unification.

In both cases the scale of unification $\Lambda$ is considered the
cutoff, and we run the theory below it. We assume, therefore, that perturbation
theory is valid. There appears a hierarchy problem. From the point of view of the spectral action this implies a
rather strange (though admissible) cutoff function. From a phenomenological point of view
this entails either unnaturally large dimensionless quantities, or
the presence of a new intermediate scale, $\Gamma$. The latter option is, of course, more desirable and we will discuss it below.

The paper is organized as follows: in section~\ref{SMRCC}, we present the action of Standard Model (SM) of particle physics,
and the standard running coupling constants. We show how the spectral action approach $-$ whose principles are summarized in the appendix~\ref{appspectralaction} $-$ fits the SM. In section~\ref{RGEs} we give the new renormalization group equations at one loop, due to the dimensions-6 operators; then, we show how these new operators affect the SM phenomenology. In section~\ref{RGF} we run the renormalization group equations to study the new coupling constants behavior, checking the possibility to improve the gauge unification point.  A final section contains conclusions and some comments and open questions.

\section{Standard Model Running Coupling Constans\label{SMRCC}}

The standard model action (including right handed neutrinos) is:
\begin{eqnarray}
\mathcal{L}_{SM} & = & -\frac{1}{4}V_{\mu\nu}^{A}V^{A\mu\nu}-\frac{1}{4}W_{\mu\nu}^{I}W^{I\mu\nu}-\frac{1}{4}B_{\mu\nu}B^{\mu\nu}+\left(D_{\mu}H\right)^{\dagger}D^{\mu}H-\frac{m^{2}}{2}H^{\dagger}H-\lambda \left(H^{\dagger}H\right)^{2}+\nonumber \\
 &  & -\left(\overline{e}\mathbf{F}_{L}H^{\dagger}l+\overline{d}\mathbf{F}_{D}H^{\dagger}q+\overline{u}\mathbf{F}_{U}H^{c\dagger}q+h.c.\right)+\mbox{Majorana mass terms} \label{eq: SM Lagrangian}
\end{eqnarray}
where $B_{\mu\nu}$, $W_{\mu\nu}^{I}$ and $V_{\mu\nu}^{A}$ are respectively
the field strengths associated with the gauge groups $U(1),\, SU(2)$
and $SU(3)$; the gauge covariant derivative is $D_{\mu}=\partial_{\mu}+ig_{3}T^{A}A_{\mu}^{A}+ig_{2}t^{I}W_{\mu}^{I}+ig_{1}\mbox{y}B_{\mu}$,
where $T^{A}$ are the $SU(3)$ generators, $t^{I}=\tau^{I}/2$ are
the $SU(2)$ generators, and y is the $U(1)$ hypercharge generator.
$H$ is the Higgs field, a $SU(2)$ scalar doublet with hypercharge $1/2$ and its
charged conjugated field defined as $H^{c}\equiv i\tau_{2}H^{*}$.
The three families of fermions are grouped together so that $\mathbf{F}_{L}$,
$\mathbf{F}_{U}$, $\mathbf{F}_{D}$ are the $3\times3$ complex Yukawa
matrices acting on the hidden flavor index of every fermion field. Since
these matrices are dominated by the Yukawa coupling $y_t$ of the top quark,
in the following we will consider this parameter only. Likewise we will
consider a single mass term for the Majorana masses: $y_\nu$.

This Lagrangian can be obtained from first principles using the spectral
action~\cite{AC2M2, NCPart1}, which is a regularized trace, with $\Lambda$ appearing as the
cutoff. We give the details of the spectral action calculations in
the appendix. For the economy of this paper the relevant part is the
fact that the spectral action requires the coupling constants of the
three gauge groups to be equal at a scale $\Lambda$, which is also
the cutoff of the theory. There is no need for a unified gauge group
at the scale $\Lambda$, which in fact may signify a phase transition
to a pre geometric phase~\cite{Kuliva}, although larger symmetries
are also possible~\cite{coldplay,CCvS}. As explained in
the appendix, the spectral action is an expansion in inverse powers
of $\Lambda^{2}$, and it enables the presence of a set of new dimension
six operators. Dimension five operators, which violate lepton number,
and do not change the properties of the Higgs boson are not present
in the expansion. The spectral action also gives relations among the
coefficients of the required dimension six operators, which are described
in detail in the appendix~\ref{appspectralaction}. The reader interested only
in the phenomenological aspect of this paper may skip the appendix, and accept
our choice of operator as a convenient one.

A complete classification of the dimension-six operators in the standard
model is given in \cite{Dim6Classification}. There it is
shown that there are 59 independent operators, preserving baryon number,
after eliminating redundant operators using the equations of motion.
Here we consider only the following dimension-six operators,
mixing the gauge field strength and the Higgs field. They are
the ones coming from the spectral action expansion:
\begin{eqnarray}
\mathcal{L}^{(6)} & = & C_{HB}H\overline{H\,}B_{\mu\nu}B^{\mu\nu}+C_{HW}H\overline{H}\,\mathbf{W}_{\mu\nu}\mathbf{W}^{\mu\nu}+C_{HV}H\overline{H}\,\mathbf{V}_{\mu\nu}\mathbf{V}^{\mu\nu}+\nonumber \\
 &  & +C_{W}\mathbf{W}_{\mu\nu}\mathbf{W}^{\nu\alpha}\mathbf{W}_{\alpha}^{\mu}+C_{V}\mathbf{V}_{\mu\nu}\mathbf{V}^{\nu\alpha}\mathbf{V}_{\alpha}^{\mu}+C_{H}\left(\overline{H}H\right)^{3}\label{dimsixlagrangian}
\end{eqnarray}
The coefficients $C$ have the dimension of an inverse energy square. The spectral action fixes their value
at the cutoff $\Lambda$. To these terms we have to add a coupling between the Higgs, the $W$ and the $B$ which is absent in the spectral action at scale $\Lambda$, but is dynamically created. With the couplings considered here no other term is induced.

\begin{figure}[htb]
\centering{}\includegraphics[width=9cm,height=7cm,keepaspectratio]{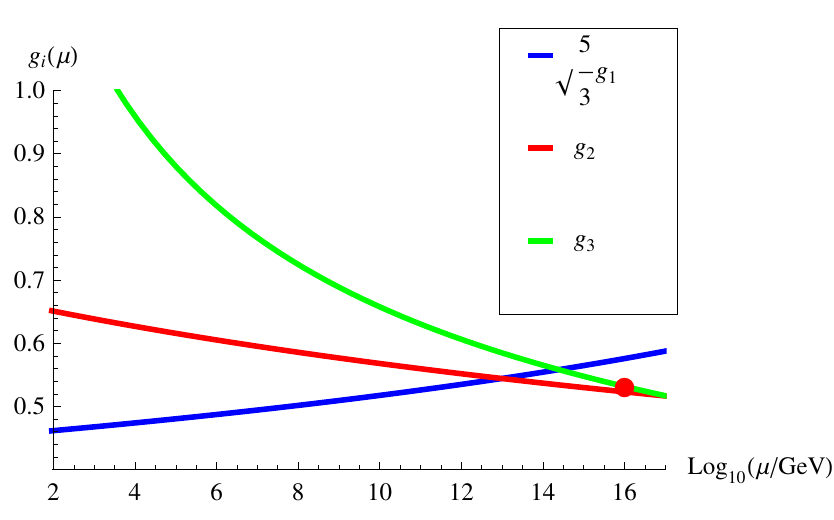}\includegraphics[width=9cm,height=7cm,keepaspectratio]{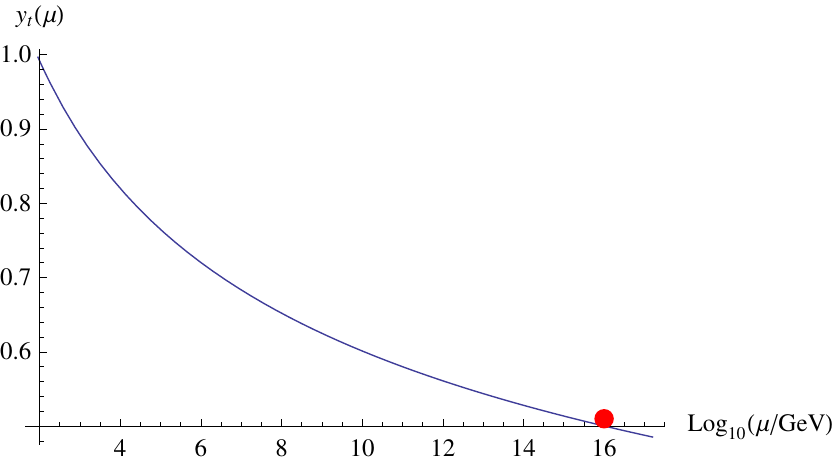}\caption{\textsl{The standard model running for the gauge coupling constant (left)
and the Yukawa coupling (right) in the spectral action approach for $(g,\rho,\Lambda)=(0.530,\,1.25,\,10^{16}GeV)$. The red dot indicates the starting value of the parameters (Log$_{10} (\frac{\Lambda}{GeV}),g$) and  (Log$_{10} (\frac{\Lambda}{GeV}),\lambda(\Lambda)$).
\label{fig:Standard gi}}}
\end{figure}

The SM running coupling constants at one loop, associated to (\ref{eq: SM Lagrangian}),
are ruled by the following equations, where we defined the dot derivations
as $16\pi^{2}\mu\frac{d}{d\mu}$:
\begin{eqnarray}
\dot{g}_{i} & = & \left(b_{i}g_{i}^{3}\right)\,\,\mbox{with}\,\left(b_{1},b_{2},b_{3}=\frac{41}{6},-\frac{19}{6},-7\right)\nonumber \\
\dot{\lambda} & = & \left(24\lambda^{2}-\left(3g_{1}^{2}+9g_{2}^{2}\right)\lambda+\frac{3}{8}\left(g_{1}^{4}+2g_{1}^{2}g_{2}^{2}+3g_{2}^{4}\right)+\left(12y_t^{2}+4y_\nu^{2}\right)\lambda-6y_t^{4}-2y_\nu^{4}\right)\nonumber \\
\dot{y}_{t} & = & \left(\frac{9}{2}y_t^{2}+y_\nu^{2}-\frac{17}{12}g_{1}^{2}-\frac{9}{4}g_{2}^{2}-8g_{3}^{2}\right)\nonumber \\
\dot{y}_{\nu} & = & \left(\frac{5}{2}y_\nu^{2}+3y_t^{2}-\frac{3}{4}g_{1}^{2}-\frac{9}{4}g_{2}^{2}\right)\label{eq: SM RGE}
\end{eqnarray}
For the purposes of this paper one loop is sufficient, the running
up to three loops can be found in~\cite{Machacek1,Machacek2,Machacek3,3loops} and references therein.
In the present case, one separately solves the equations for the gauge coupling
constants and the other couplings; for the former, the boundary conditions
are given at the electro-weak scale by the experimental values \cite{particledata},
\begin{eqnarray}
g_{1}(m_{Z}) = 0.358,\,\, g_{2}(m_{Z})=0.651,\,\, g_{3}(m_{Z})=1.221\,\label{eq:InitialCondition}
\end{eqnarray}
while for the other coupling constants $\lambda,y_t$ and $y_\nu$
the boundary conditions  are taken at the cut-off scale $\Lambda$
that is the scale at which the spectral action lives. These boundary
conditions use the parameters of the fermions that are the inputs
in the Dirac operator~\eqn{Diracoperator}, as shown in the appendix~\ref{appspectralaction},
\begin{align}
\lambda(\Lambda) & =\frac{4\left(\rho^2+3\right)}{\left(\rho+3\right)^{2}}g^{2}\nonumber \\
y_t(\Lambda) & =\sqrt{\frac{4}{\rho^2+3}}g\nonumber \\
y_\nu(\Lambda) & =\sqrt{\frac{4\rho^2}{\rho^2+3}}g\label{eq:Boundary conditions}
\end{align}
$\rho$ is a free parameter such that $y_\nu(\Lambda)=\rho\,y_t(\Lambda)$
and $g\equiv g_{3}(\Lambda)=g_{2}(\Lambda)=\frac{5}{3}g_{1}(\Lambda)$.  Since the coupling
constants $g_{i}$ do not meet exactly, forming a triangle, one takes for $\Lambda$ a
range of values beetwen the extremal points of the triangle.  The results, for a particular set of values of the parameters \textsl{$(g,\rho,\Lambda)=(0.530,\,1.25,10^{16}GeV)$,}
are plotted in Fig.~\ref{fig:Standard gi}.

\begin{figure}
\centering{}\includegraphics[width=9cm,height=7cm,keepaspectratio]{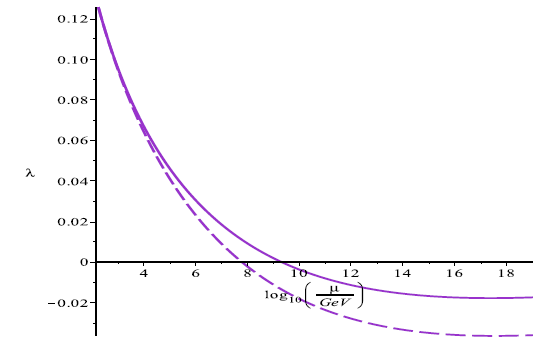}\includegraphics[width=9cm,height=7cm,keepaspectratio]{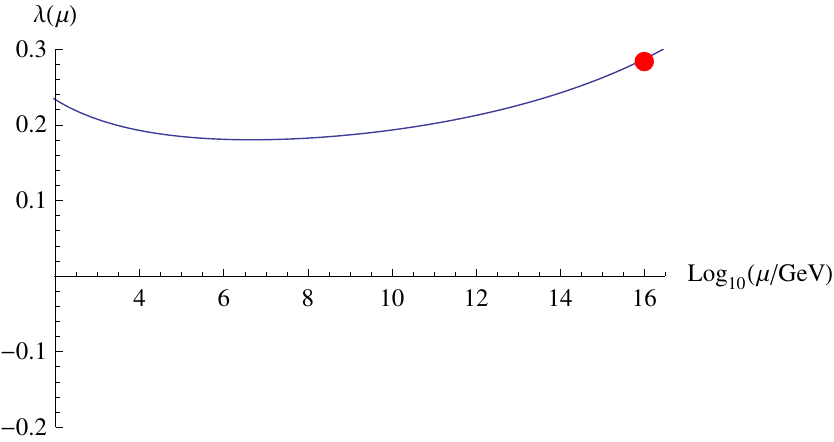}\caption{\textsl{On the left, the standard model running for the coupling constant $\lambda$
starting from $\lambda(m_{Z})=0.130$ corresponding to $M_{H}=125GeV$.
The dashed and solid lines represent the one and two loop respectively. On the right, the $\lambda$ behaviour starting from the red point of the spectral action and culminating in the prediction $\lambda(m_{Z})=0.240$.   }
\label{lambdabehaviours}}
\end{figure}

After running these couplings from unification energy $\Lambda$ to low energy $M_Z$, we compare the values of $y_t(M_Z)$ and $\lambda(M_Z)$ with their experimental values
\begin{eqnarray}
y^{exp}_t(M_Z)=0.997,\,\,\ & \lambda^{exp}(M_Z)=0.130 \label{expvalues}
\end{eqnarray}
In Fig.~\ref{fig:Standard gi} we can see the good agreement between $y_t(M_Z)$  predicted by the spectral action and its experimental value. Very different is  the case for the Higgs self-coupling $\lambda$, Fig.~\ref{lambdabehaviours}, whose predicted value, in the spectral action approach, is around 0.240 with a resulting Higgs mass $M_H=\sqrt{2 \lambda v^2 }\simeq 170$~ GeV. On the other hand, the experimental value for the Higgs mass ($\simeq$125~GeV) leads to
the instability problem for the self-interaction parameter $\lambda$,
which becomes negative at a scale of the order of $10^{8}GeV$; two
loop calculations make the situation slightly worse as one can see on the left side of Fig.~\ref{lambdabehaviours}.

 A negative $\lambda$ means an instability and renders the
model inconsistent, although it may just mean the presence of a long lived metastable state~\cite{metastable1,metastable2}. The spectral action model can be fixed~\cite{Stephan,coldplay,resilience,CCvS,BoyleFarnsworthsigma}
with the introduction of a scalar field, $\sigma$, possibly coming
from a larger symmetry, connected with the fluctuations of a Majorana
neutrino mass term in the action. Since the running of the Higgs parameters
do not affect strongly the running of the coupling constants (which
are the true aim of this paper), nor does this field $\sigma$, we
will not consider it in what follows. However, a more complete and accurate analysis will necessitate also this element, and it is in progress. Also the presence of gravitational couplings in the spectral action could alter significativly the running at high energy leading to an asymptotically free theory at the Planck scale~\cite{Devastato}.

\section{Coupling Constants RGEs\label{RGEs}}
\setcounter{equation}{0}

In this section we give the new renormalization group equations (RGEs)
at one loop due to the dimensions-6 operators in the Lagrangian~\eqn{dimsixlagrangian}.
Although the choice of the dimension six operators and some of characteristics
of the Lagrangian are coming from the spectral action, this section
can be read independently of it.

The full one-loop contributions to the SM running for dimension six
operators have been calculated in~\cite{Grojean,Manohar1,Manohar2,JenkinsIII}. The
modifications to the standard model RGEs are given by the following
new terms to be added to the rhs of~\eqn{eq: SM RGE}:
\begin{eqnarray}
\delta\dot{g_{3}} & = & -4m_{H}^{2}g_{3}C_{HV}\nonumber \\
\delta\dot{g_{2}} & = & -4m_{H}^{2}g_{2}C_{HW}\nonumber \\
\delta\dot{g_{1}} & = & -4m_{H}^{2}g_{1}C_{HB}\nonumber \\
\delta\dot{\lambda}\  & = & m_{H}^{2}\left(9g_{2}^{2}C_{HW}+3g_{1}^{2}C_{HB}+12C_{H}+3g_{1}g_{2}C_{HWB}\right)\nonumber \\
\delta\dot{y}_{t,\nu} & = & 0\label{eq:modification beta}
\end{eqnarray}
and the RGEs  for the dim-6 coupling constants are given by
\begin{align}
\dot{C}_{HB} & =C_{HB}\left(12\lambda+2\left(3y_t^{2}+y_\nu^{2}\right)+\frac{85}{6}g_{1}^{2}-\frac{9}{2}g_{2}^{2}\right)+3C_{HWB}g_{1}g_{2}\nonumber \\
\dot{C}_{HW} & =C_{HW}\left(12\lambda+2\left(3y_t^{2}+y_\nu^{2}\right)-\frac{47}{6}g_{1}^{2}-\frac{5}{2}g_{2}^{2}\right)+C_{HWB}g_{1}g_{2}-15C_{W}g_{2}^{3}\nonumber \\
\dot{C}_{HV} & =C_{HV}\left(12\lambda+2\left(3y_t^{2}+y_\nu^{2}\right)-\frac{3}{2}g_{1}^{2}-\frac{9}{2}g_{2}^{2}-14g_{3}^{2}\right)\nonumber \\
\dot{C}_{HWB} & =C_{HWB}\left(4\lambda+2\left(3y_t^{2}+y_\nu^{2}\right)+\frac{19}{3}g_{1}^{2}+\frac{4}{3}g_{2}^{2}\right)+2g_{1}g_{2}\left(C_{HB}+C_{HW}\right)+3C_{W}g_{1}g_{2}^{2}\nonumber \\
\dot{C}_{W} & =\frac{29}{2}C_{W}g_{2}^{2}\nonumber \\
\dot{C}_{V} & =15C_{V}g_{3}^{2}\nonumber \\
\dot{C}_{H} & =C_{H}\left(108\lambda+6\left(3y_t^{2}+y_\nu^{2}\right)-\frac{9}{2}g_{1}^{2}-\frac{27}{2}g_{2}^{2}\right)-3C_{B}g_{1}^{2}\left(g_{1}^{2}+g_{2}^{2}-4\lambda\right)+\nonumber \\
 & \,\,\,\,+3C_{W}g_{2}^{2}\left(12\lambda-3g_{2}^{2}-g_{1}^{2}\right)+C_{HWB}\left(12\lambda g_{1}g_{2}-3g_{1}^{3}g_{2}-3g_{1}g_{2}^{3}\right)\label{eq: RGE Correction-1}
\end{align}
Although the spectral action does not contain explicitly the term
$C_{HWB}H^{2}W_{\mu\nu}B^{\mu\nu}$, due to the unimodular condition,
the coupling constant $C_{HWB}$ is however induced by the running
of $C_{HB},\, C_{HW}$ and $C_{W}$.

In the framework of the spectral action these equations are solved
with boundary conditions at the cut-off scale $\Lambda$ given by
the coefficients appearing in (\ref{eq:A6 Expression}):
\begin{align}
C_{HB}(\Lambda) & =-\frac{f_{6}}{16\pi^{2}\Lambda^{2}}\frac{4\left(3\rho^2+17\right)}{9\left(\rho^2+3\right)}g^{4}\,\,,\, C_{HW}(\Lambda)=-\frac{f_{6}}{16\pi^{2}\Lambda^{2}}\frac{4}{3}g^{4}\,\,,\, C_{HV}(\Lambda)=-\frac{f_{6}}{16\pi^{2}\Lambda^{2}}\frac{16}{3\left(\rho^2+3\right)}g^{4}\,\,,\nonumber \\
C_{H}(\Lambda) & =-\frac{f_{6}}{16\pi^{2}\Lambda^{2}}\frac{512(\rho^{6}+3)}{3\left(\rho^2+3\right)^{3}}g^{6}\,\,,\,\,\,\,\, C_{W}(\Lambda)=-\frac{f_{6}}{16\pi^{2}\Lambda^{2}}\frac{26}{15}g^{3}\,\,,\,\,\,\,\, C_{V}(\Lambda)=-\frac{f_{6}}{16\pi^{2}\Lambda^{2}}\frac{26}{15}g^{3}\,.\,\,\,\label{eq: constrains new}
\end{align}
The coupling $C_{HWB}$ is set to zero at the cut-off scale $C_{HWB}(\Lambda)=0$
since it does not appear in the spectral action.

In (\ref{eq: constrains new}) $g\equiv g_{3}(\Lambda)=g_{2}(\Lambda)=\frac{5}{3}g_{1}(\Lambda)$
is the value of the gauge coupling constants at the cut-off scale
which, therefore, is identified with the unification scale. These two
constants, $g$ and $\Lambda$, together with the ratio $\rho$ and
the parameter $f_{6}$ appearing in the spectral action, will be the
four free parameters of this model.

There are also constraints at low energy to satisfy. The values of
the $g_{i}$'s are known at the scale of the top mass with very high
precision, and the parameters $\lambda$ and the $y_t$ are related to
the Higgs and top mass. As we said earlier, the spectral action requires a positive value of $\lambda$ at the cutoff scale $\Lambda$, (\ref{eq:Boundary conditions}), and without
the field $\sigma$, it predicts a mass of the Higgs at 170~GeV. However, the presence of higher-order operators in the action alters the form of the usual coupling constants, leading to a new phenomenology which we outline in the following section.

\subsection{New phenomenology}

In this section, following~\cite[sect.5]{JenkinsIII} we give the main modifications to the SM phenomenology
due to the dim-6 Lagrangian, i.e. the new form of the observables
measured at the electroweak scale.
The new operators, in fact, alter the definition of the SM parameters at tree level in several
ways.

First of all, we focus on the effects of the dimension-six Lagrangian
on the Higgs mass $m_{H}$ and the self-coupling $\lambda$. The dim-6
operator $C_{H}\bar{H}H$ changes the shape of the scalar doublet
potential at order $C_{H}v^{2}$ to
\begin{equation}
V(H)=-\frac{m^{2}}{2}H^{\dagger}H+\lambda\left(H^{\dagger}H\right)^{2}-C_{H}\left(H^{\dagger}H\right)^{3}\label{eq:new potential}
\end{equation}
generating the new minimum
\begin{eqnarray}
\left\langle H^{\dagger}H\right\rangle  & = & \frac{1}{3C_{H}}\left(\lambda-\sqrt{\lambda^{2}-3C_{H}\lambda v^{2}}\right)\nonumber \\
 & \simeq & \frac{v^{2}}{2}\left(1+\frac{3C_{H}v^{2}}{4\lambda}\right)\equiv\frac{v_{T}^{2}}{2}
\end{eqnarray}
in the second line we have expanded the exact solution to first order
in $C_{H}.$ Therefore the shift in the vacuum expectation value is
proportional to $C_{H}v^{2},$ which is of order $f_{6}\frac{v^{2}}{\Lambda^{2}}.$
On expanding the potential~(\ref{eq:new potential}) around the minimum and neglecting kinetics corrections,
\begin{equation}
H=\frac{1}{\sqrt{2}}\left(\begin{array}{c}
0\\
h+v_{T}
\end{array}\right),
\end{equation}
we find for the Higgs boson mass
\begin{equation}
m_{H}^{2}=2\lambda v_{T}^{2}\left(1-\frac{3C_{H}v^{2}}{2\lambda}\right)\label{Higgsmass}
\end{equation}
At the same time the gauge fields and the gauge couplings are also
affected by the dim-6 couplings.

In the broken theory the $X^{2}H^{2}$ operators (with $X$ being any field strength) contribute to the
gauge kinetic energies, through the Lagrangian terms
\begin{eqnarray}
\left(\mathcal{L}_{SM}+\mathcal{L}_{6}\right)_{kin} & = & -\frac{1}{4}B_{\mu\nu}B^{\mu\nu}-\frac{1}{4}G_{\mu\nu}G^{\mu\nu}-\frac{1}{4}W_{\mu\nu}^{3}W_{3}^{\mu\nu}-\frac{1}{2}W_{\mu\nu}^{+}W_{-}^{\mu\nu}+ \\
 &  & +\frac{1}{2}v_{T}^{2}\left(C_{HB}B_{\mu\nu}B^{\mu\nu}+C_{HW}W_{\mu\nu}W^{\mu\nu}+C_{HG}G_{\mu\nu}G^{\mu\nu}-C_{HWB}W_{\mu\nu}^{3}B^{\mu\nu}\right) \nonumber
\end{eqnarray}
while for the mass terms of the gauge bosons, arising from $(D_{\mu}H)^{\dagger}(D^{\mu}H$),
we have
\begin{equation}
\left(\mathcal{L}_{SM}+\mathcal{L}_{6}\right)_{mass}=\frac{1}{4}g_{2}^{2}v_{T}^{2}W_{\mu\nu}^{+}W_{-}^{\mu\nu}+\frac{1}{8}v_{T}^{2}\left(g_{2}W_{\mu}^{3}-g_{1}B_{\mu}\right)^{2}+\frac{1}{16}v_{T}^{4}C_{HD}\left(g_{2}W_{\mu}^{3}-g_{1}B_{\mu}\right)^{2}
\end{equation}
The gauge fields have to be redefined, so that the kinetic terms are
properly normalized and diagonal,
\begin{equation}
G_{\mu}=\mathcal{G}_{\mu}\left(1+C_{HG}v_{T}^{2}\right),\,\, W_{\mu}=\mathcal{W}_{\mu}\left(1+C_{HW}v_{T}^{2}\right),\,\, B_{\mu}=\mathcal{B}_{\mu}\left(1+C_{HB}v_{T}^{2}\right),
\end{equation}
so that the modified coupling constants become
\begin{equation}
\bar{g}_{3}=g_{3}\left(1+C_{HG}v_{T}^{2}\right),\,\,\bar{g}_{2}=g_{2}\left(1+C_{HW}v_{T}^{2}\right),\,\,\bar{g}_{1}=g_{1}\left(1+C_{HB}v_{T}^{2}\right),
\end{equation}
and the products $g_{1}B_{\mu}=\bar{g}_{1}\mathcal{B}_{\mu}$ etc. are unchanged.
Therefore, the electroweak Lagrangian is
\begin{eqnarray}
\mathcal{L} & = & -\frac{1}{4}\mathcal{B}_{\mu\nu}\mathcal{B}^{\mu\nu}-\frac{1}{4}\mathcal{W}_{\mu\nu}^{3}\mathcal{W}_{3}^{\mu\nu}-\frac{1}{2}\mathcal{W}_{\mu\nu}^{+}\mathcal{W}_{-}^{\mu\nu}-\frac{1}{2}\left(v_{T}^{2}C_{HWB}\right)\mathcal{W}_{\mu\nu}^{3}\mathcal{B}^{\mu\nu}\nonumber \\
 &  & +\frac{1}{4}\bar{g}_{2}^{2}v_{T}^{2}\mathcal{W}_{\mu\nu}^{+}\mathcal{W}_{-}^{\mu\nu}+\frac{1}{8}v_{T}^{2}\left(\bar{g}_{2}\mathcal{W}_{\mu}^{3}-\bar{g}_{1}\mathcal{B}_{\mu}\right)^{2}+\frac{1}{16}v_{T}^{4}C_{HD}\left(\bar{g}_{2}\mathcal{W}_{\mu}^{3}-\bar{g}_{1}\mathcal{B}_{\mu}\right)^{2}.
\end{eqnarray}
The mass eigenstate basis is given by,~\cite[eq.5.21]{JenkinsIII},
\begin{equation}
\left[\begin{array}{c}
\mathcal{W}_{\mu}^{3}\\
\mathcal{B}_{\mu}
\end{array}\right]=\left[\begin{array}{cc}
1 & -\frac{1}{2}v_{T}^{2}C_{HWB}\\
-\frac{1}{2}v_{T}^{2}C_{HWB} & 1
\end{array}\right]\left[\begin{array}{cc}
\mbox{cos}\bar{\theta} & \mbox{sin}\bar{\theta}\\
\mbox{-sin}\bar{\theta} & \mbox{cos}\bar{\theta}
\end{array}\right]\left[\begin{array}{c}
\mathcal{Z}_{\mu}^{3}\\
\mathcal{A}_{\mu}
\end{array}\right],
\end{equation}
with $\bar{\theta}$, rotation angle, given by
\begin{equation}
\mbox{tan}\bar{\theta}=\frac{\bar{g}_{1}}{\bar{g}_{2}}+\frac{v_{T}^{2}}{2}C_{HWB}\left[1-\frac{\bar{g}_{1}^{2}}{\bar{g}_{2}^{2}}\right]\,.
\end{equation}
The photon remains massless and the $W$ and $Z$ masses are
\begin{eqnarray}
M_{W}^{2} & = & \frac{\bar{g}_{2}^{2}v_{T}^{2}}{4},\nonumber \\
M_{Z}^{2} & = & \frac{(\bar{g}_{1}^{2}+\bar{g}_{2}^{2})v_{T}^{2}}{4}+\frac{1}{8}v_{T}^{4}C_{HD}\left(\bar{g}_{1}^{2}+\bar{g}_{2}^{2}\right)+\frac{1}{2}v_{T}^{4}\bar{g}_{1}\bar{g}_{2}C_{HWB}\label{eq: new gauge masses}
\end{eqnarray}
The covariant derivative has the form
\begin{equation}
D_{\mu}=\partial_{\mu}+i\frac{\bar{g}_{2}}{\sqrt{2}}\left[\mathcal{W}_{\mu}^{+}T^{+}+\mathcal{W}_{\mu}^{-}T^{-}\right]+ig_{Z}\left[T_{3}-\mbox{sin}\bar{\theta}^{2}Q\right]\mathcal{Z}_{\mu}+i\bar{e}Q\mathcal{A}_{\mu},
\end{equation}
where $Q=T_{3}+Y$ and the effective couplings become,
\begin{eqnarray}
\bar{e} & = & \frac{\bar{g}_{1}\bar{g}_{2}}{\sqrt{\bar{g}_{1}^{2}+\bar{g}_{2}^{2}}}\left[1-\frac{\bar{g}_{1}\bar{g}_{2}}{\bar{g}_{1}^{2}+\bar{g}_{2}^{2}}v_{T}^{2}C_{HWB}\right]=\bar{g}_{2}\mbox{sin}\bar{\theta}-\frac{1}{2}\mbox{cos}\bar{\theta}\bar{g}_{2}v_{T}^{2}C_{HWB}\,,\nonumber \\
\bar{g}_{Z} & = & \sqrt{\bar{g}_{1}^{2}+\bar{g}_{2}^{2}}+\frac{\bar{g}_{1}\bar{g}_{2}}{\sqrt{\bar{g}_{1}^{2}+\bar{g}_{2}^{2}}}v_{T}^{2}C_{HWB}=\frac{\bar{e}}{\mbox{sin}\bar{\theta}\mbox{cos}\bar{\theta}}\left[1+\frac{\bar{g}_{1}^{2}+\bar{g}_{2}^{2}}{2\bar{g}_{1}\bar{g}_{2}}v_{T}^{2}C_{HWB}\right]\,,\nonumber \\
\mbox{sin}\bar{\theta}^{2} & = & \frac{\bar{g}_{1}^{2}}{\bar{g}_{1}^{2}+\bar{g}_{2}^{2}}+\frac{\bar{g}_{1}\bar{g}_{2}\left(\bar{g}_{2}^{2}-\bar{g}_{1}^{2}\right)}{\bar{g}_{1}^{2}+\bar{g}_{2}^{2}}v_{T}^{2}C_{HWB}\,.\label{eq: new effective couplings}
\end{eqnarray}
Considering (\ref{eq: new effective couplings}) and (\ref{eq: new gauge masses}),
the experimental values for the $W$ and $Z$ masses and couplings
fix $\bar{g}_{1},$$\bar{g}_{2},$$v_{T}$, $C_{HWB}$ and $C_{HD}.$
This procedure consists of solving 5 equations in 5 variables:
the unique solution of this system is given by the classical values
for $\bar{g}_{1},$$\bar{g}_{2}$ and $v_{T}$, i.e.
\begin{equation}
\bar{g}_{1}=0.358,\,\,\bar{g}_{2}=0.651,\,\, v_{T}=246\, GeV
\end{equation}
while the dim-6 parameters $C_{HWB}$ and $C_{HD}$ have to give
negligble corrections to the standard results. This means the products
$v_{T}^{2}C_{HWB}$ and $v_{T}^{2}C_{HD}$ have to be, at least, of the order $10^{-3}$, i.e. $C_{HWB,D}\lesssim10^{-7}GeV^{-2}$.

\section{Running of the constants\label{RGF}}
\setcounter{equation}{0}
In the following section we run the renormalization group equations, presented in sect.~\ref{RGEs}, to study the modification of the coupling constants behavior, due to the dim-6 operators. We check the possibility, for these new terms, to give a gauge unification point and to return values for the coupling constants compatible with the spectral action predictions.

\subsection{Renormalization group flow}

One can run the equations of the renormalization group in two directions. A ``bottom-up'' running assumes boundary values for the various constants at low energy (usually the $Z$ or top mass) and runs toward higher energies. This is the way Fig.~\ref{fig:Standard gi}
has been obtained.
On the contrary the spectral action is defined at the high energy scale $\Lambda$, and its strength lies in the fact that it specifies the boundary conditions of all constants there. Therefore a ``top-down'' approach is more natural. In this paper we follow a combined approach.

We start at the scale $\Lambda$ in the range $10^{13-17}$~GeV. At this energy we give the boundary values given by the spectral action. In particular we use for the dimension six terms the values we have calculated and presented in~\eqn{eq: constrains new}. The top-down running depends on four other parameters (described below) and gives a set of values for all of physical parameters at low energy. The parameters we find are not too distinct from the experimentally known ones, but there are discrepancies. As it should be: the heat kernel expansion is akin to a one loop calculation and, apart form any other incomplete aspect of the theory, it would be unreasonable to find the correct values for all parameters. The values one finds are however close to the experimental ones for the three $g_i$ and $y_t$, while as remarked earlier $\lambda$, which is the parameter appearing in the Higgs mass, is off by nearly a factor two. The top-down running gives a set of values of the dimension six couplings $C_i$ at $M_Z$ .

We then performed a bottom-up running to see if the presence of the new terms could give a unification point, and we found that in several cases it does. As boundary conditions we used the experimental values for the $g_i$'s and $y_t$ and the low energy values of the $C_i$'s obtained in the top-down running. The case of $\lambda$ deserves  a little discussion. Since the experimental and spectral action values are quite different, the qualitative behaviour in the two cases are different. On the other side, it is known that the problem is fixed by the presence of another field ($\sigma$), which we do not discuss in this paper. We have therefore performed our analysis in the two cases, i.e.\ the value of $\lambda$ obtained by the spectral action, and the experimental one. The strategy we followed is synthesized in Table~\ref{scheme}.
\begin{table}[htb]
\begin{equation}
\left[\begin{array}{ccc}
\Lambda\ \mbox{scale}\overset{\mbox{\small RGEs}}{\longrightarrow}M_{Z}\,\mbox{scale} &  & M_{Z}\,\mbox{scale}\overset{\mbox{\small RGEs}}{\longrightarrow}\Lambda\mbox{scale}\\
\mbox{\small In:}\{\mbox{\small eq.(\ref{eq: constrains new})}\} &  & \mbox{\small In:}\{C_{i}(M_{Z}),\, g_{i}^{exp}\}\\
\mbox{\small Out:}\{\mbox{\small}C_{i}(M_{Z})\} &  & \mbox{\small Out:}\{C_{i}(\Lambda),g_{i}(\Lambda)\}\\
\underbrace{\,\,\,\,\,\,\,\,\,\,\,\,\,\,\,\,\,\,\,\,\,\,\,\,\,\,\,\,\,\,\,\,\,\,\,\,\,\,\,\,\,\,\,\,\,\,\,\,\,\,\,\,\,\,\,\,\,\,} &  & \underbrace{\,\,\,\,\,\,\,\,\,\,\,\,\,\,\,\,\,\,\,\,\,\,\,\,\,\,\,\,\,\,\,\,\,\,\,\,\,\,\,\,\,\,\,\,\,\,\,\,\,\,\,\,\,\,\,\,\,\,}\\
{\color{blue}\mbox{top-down\,\ running}} &  & {\color{blue}\mbox{bottom-up\,\ running}}
\end{array}\right]
\end{equation}
\caption{\textsl{\small Resolution scheme adopted for the renormalization group
flow. Varying $\Lambda,\, g,\,\rho$ and $f_{6}$ we solve the RGEs,
starting from the unification scale $\Lambda$ down to the $M_{Z}$
scale, and we use the resulting values for the dim-6 parameters together
with the experimental values for the usual dim-4 couplings $g_{i}^{exp},\,\lambda^{exp},\, y_{top}^{exp}$
to run again toward high energies.}}
\label{scheme}
\end{table}

The second case, in which we used the experimental values as initial
conditions, can be considered on a purely phenomenological basis, to show that higher dimension operators may cause unifications of the constants at one loop.

\subsection{Top-Down running}

In the spectral action model we have four free parameters: the value
of the gauge coupling constants at the unification, $g$. The value
of the cut-off and unification scale, $\Lambda$. The ratio between
top and neutrino Yukawa couplings, $\rho$. The momentum $f_{6}$
which will fix the new physics scale $\Gamma$. This last parameter appears as coefficient to the dimension six operators with the combination $f_6/\Lambda^2$, and therefore effectively defines a new energy scale.

All parameters have a particular range in which we expect they
could be chosen. From the SM running of the
gauge coupling constants we know $g$ is expected around $0.55\pm0.03$,
while $\Lambda$ has a more significant range between $10^{13}$~GeV
and $10^{17}$~GeV. The ratio $\rho$ between the top and neutrino
Yukawa couplings should be expected of $O(1)$. The value of the parameter
$f_{6}$ requires a separate discussion. From the internal logic of
the spectral action its ``natural'' value would be of order unity,
or not much larger. Such a value would however make the corrections
to the running totally irrelevant. The parameter appears with a denominator
in $\Lambda^{2}$, and the corrections are often quadratic in this
ratio. On the other side, from the phenomenology of electroweak processes
it can be expected the effects of these new physics terms on the measured
signal strength for $H\rightarrow\mathrm{\gamma\gamma}$ decay, whose
measured value is given by ATLAS and CMS~\cite{Atlas,CMS}. To obtain comparable
data the new physics scale has to be fixed around $\Gamma\thicksim1-10$~TeV.
This leads to expected values for the dim-6 coefficients $C_{i}$
around $10^{-6}-10^{-8}\mbox{GeV}^{-2}$ . The range for $f_{6}$
will be $\thicksim\Lambda^{2}/\Gamma^{2}$, i.e.\ $10^{20-28}$. Given
the fact that the cutoff function is undetermined in the scheme, such
numbers are allowed, although a more physical explanation of their
size would be preferable. The spectral action, given by an expansion
valid below the unification scale, gives a framework to use a perturbative
expansion valid beyond the scale of new physics, although it does not explain it. From the spectral point of view this is a weak point, the presence of such a high value for $f_6$ is very strange and creates an unnatural hierarchy with the other coefficients.

Since the point of the calculation was to verify the possibility of unification, the top-down calculation has been performed with the aim of obtaining values which would be a good starting point for the bottom up calculation. We did search for the best solutions for the range of parameters above. We performed first a coarse search to restrict the range, and then optimized the input parameters to find a good unification point. For the scope of this paper, i.e.\ to show that dimension six operators could give unification, this is sufficient.

The boundary conditions at $M_Z$ for the subsequent bottom-up run approach are the experimental values for the $g_i$ and $y_t$, and the values obtained from the top-down for the $C_i$'s. In the case of $\lambda$ we have the two choices: either the values obtained from the top down, or the one from experiment. Since these two are different, in the following we present both cases.

\subsubsection{Spectral action value for $\lambda$}

In the following table we describe the values of the free parameters we used which will enable the best unification.

Table~\ref{topdownvariation} shows, for various values of $\Lambda$, the parameters used for the top-down running, and the value of the couplings at low energy, shown  as ratio with respect to the experimental value, corrected as described in the previous section: $\gamma_i=\frac{g_i(M_Z)}{\bar g^{exp}_i}$ and $\gamma_t=\frac{y_t(M_Z)}{y_t^{exp}}$. The values for $\lambda$ are not shown since, for the reasons described above, they are not significant.

\begin{table}[htp]
\begin{center}
\begin{tabular}{||c|c|c|c|c|c|c|c||}
\hline
$\Lambda$~GeV  &  $g(\Lambda)$ & $\rho(\Lambda)$  &$\frac{f_6}{16\pi^2\Lambda^2}\mbox{Gev}^{-2}$ & $ \gamma_1$  &  $ \gamma_2$ &  $ \gamma_3$ &  $ \gamma_t$  \\
\hline
$10^{14}$ &  0.580 & 1.6  &$4.8~10^{-6}$ &$ 1.0 $ & $ 1.0 $  &  $1.0 $ &  $1.0 $   \\
\hline
$10^{15}$ &  0.570 & 1.9  &$7.3~10^{-6}$ & $0.98 $  &  $1.0 $ &  $1.0 $ &  $1.0$  \\
\hline
$10^{16}$ &  0.550 & 1.9  &$6.9~10^{-6}$ & $0.95 $  &  $0.99 $ &  $1.0$ &  $1.0$  \\
\hline
$10^{17}$ &  0.540 & 2.0  &$8.3~10^{-6}$ & $0.93$  &  $0.97 $ &  $1.1 $ &  $1.0$  \\
\hline
\end{tabular}\caption{\sl The values of the coupling constants at $M_Z$ compared with the experimental values for the top-down running. The values of the free parameters are optimized for the subsequent bottom-up run. \label{topdownvariation} }
\end{center}
\end{table}

Note that the choice of parameters has been made to optimize the subsequent bottom-up running. The amount of variations with respect to the experimental values for the couplings could be made smaller with a different choice of $g, f_6$ and $\rho$. This top-down running gives values for the $C_i$'s, which are shown in Table~\ref{topdownCi}.
\begin{table}[htp]
\begin{center}
\begin{tabular}{||c|c|c|c|c|c|c|c||}
\hline
%$\Lambda$~GeV  &  $C_{HWB}\mbox{GeV}^{-2}$ &  $C_{W}\mbox{GeV}^{-2}$ &  $C_{V}\mbox{GeV}^{-2}$ &  $C_{HV}\mbox{GeV}^{-2}$ &  $C_{H}\mbox{GeV}^{-2}$ &  $C_{HB}\mbox{GeV}^{-2}$ &  $C_{HW}\mbox{GeV}^{-2}$  \\
$\Lambda$  &  $C_{HWB}$ &  $C_{W}$ &  $C_{V}$ &  $C_{HV}$ &  $C_{H}$ &  $C_{HB}$ &  $C_{HW}$  \\
\hline
$10^{14}$ &  $1.1\,10^{-7}$ & $-5.8\,10^{-7}$  & $-2.7\,10^{-7}$ & $-1.1\,10^{-6}$  &  $3.8\,10^{-8}$ &  $-1.7\,10^{-7}$ &  $-7.5\,10^{-7}$  \\
\hline
$10^{15}$ &  $1.4\,10^{-7}$ & $-8.1\,10^{-7}$  & $-3.3\,10^{-7}$ & $-1.4\,10^{-6}$  &  $5.6\,10^{-8}$ &  $-2.1\,10^{-7}$ &  $-9.9\,10^{-7}$  \\
\hline
$10^{16}$ &  $1.2\,10^{-7}$ & $-6.7\,10^{-7}$  & $-2.6\,10^{-7}$ & $-1.3\,10^{-6}$  &  $4.2\,10^{-8}$ &  $-1.7\,10^{-7}$ &  $-8.2\,10^{-7}$  \\
\hline
$10^{17}$ &  $1.3\,10^{-7}$ & $-7.4\,10^{-7}$  & $-2.5\,10^{-7}$ & $-1.4\,10^{-6}$  &  $4.6\,10^{-8}$ &  $-1.7\,10^{-7}$ &  $-8.8\,10^{-7}$  \\
\hline
\end{tabular}\caption{\sl The values of the coefficients of the dimension six operators at $M_Z$ . The values of the free parameters are the ones in Table~\ref{topdownvariation}. All $C_i$'s are in GeV${}^{-2}$.}\label{topdownCi}
\end{center}
\end{table}

One can see that with the choice of parameters, mainly $f_6$, the $C_i$'s are in the range expected by a new physics scale of the order of 1~TeV.

\subsubsection{Experimental value for $\lambda$}

The values described above are made with parameters which are natural in the framework of the spectral action, but from the phenomenological point of view, since we now have the mass of the Higgs, and therefore the value of $\lambda(M_Z)$, we can also perform the analysis using as boundary condition the experimental value. As in the previous subsection the parameters are chosen in such a way to optimize the subsequent bottom-up run.
Tables~\ref{topdownvariationexp} and~\ref{topdownCiexp} are the counterparts of~\ref{topdownvariation} and~\ref{topdownCi} for the case optimized for unification using as input the experimental value of $\lambda$ at $M_Z$. Of course some
principle like the spectral action must be operating in the background,
to make sense of the fact that we are running the theory above the
scale $\Gamma$ all the way to the unification point.

\begin{table}[htp]
\begin{center}
\begin{tabular}{||c|c|c|c|c|c|c|c||}
\hline
$\Lambda$~GeV  &  $g(\Lambda)$ & $\rho(\Lambda)$  &$\frac{f_6}{16\pi^2\Lambda^2}\mbox{Gev}^{-2}$ & $ \gamma_1$  &  $ \gamma_2$ &  $ \gamma_3$ &  $ \gamma_t$  \\
\hline
$10^{14}$ &  0.580 & 1.1  &$1.1~10^{-5}$ &$ 0.98 $ & $ 0.95 $  &  $0.80 $ &  $1.0 $   \\
\hline
$10^{15}$ &  0.560 & 0.7 &$8.3~10^{-6}$ & $0.98 $  &  $0.96 $ &  $0.85 $ &  $1.1$  \\
\hline
$10^{16}$ &  0.550 & 1.0  &$9.6~10^{-6}$ & $0.98 $  &  $0.96 $ &  $0.89$ &  $1.1$  \\
\hline
$10^{17}$ &  0.540 & 0.9  &$8.3~10^{-6}$ & $0.99$  &  $0.98 $ &  $0.95 $ &  $1.2$  \\
\hline
\end{tabular}\caption{\sl The values of the coupling constants at $M_Z$ compared with the experimental values for the top-down running. The values of the free parameters are optimized for the subsequent bottom-up run. The initial value of $\lambda(M_Z)$ is the experimental one.}\label{topdownvariationexp}
\end{center}
\end{table}

One can see that with respect to the previous case the values of the $\gamma$'s are slightly worse, showing that in this case the result of the top-down running spectral action ``predictions'' are off. This is not surprising because for the subsequent running (for which these values are optimized) the connections with the spectral action are weaker.
\begin{table}[htp]
\begin{center}
\begin{tabular}{||c|c|c|c|c|c|c|c||}
\hline
%$\Lambda$~GeV  &  $C_{HWB}\mbox{GeV}^{-2}$ &  $C_{W}\mbox{GeV}^{-2}$ &  $C_{V}\mbox{GeV}^{-2}$ &  $C_{HV}\mbox{GeV}^{-2}$ &  $C_{H}\mbox{GeV}^{-2}$ &  $C_{HB}\mbox{GeV}^{-2}$ &  $C_{HW}\mbox{GeV}^{-2}$  \\
$\Lambda$  &  $C_{HWB}$ &  $C_{W}$ &  $C_{V}$ &  $C_{HV}$ &  $C_{H}$ &  $C_{HB}$ &  $C_{HW}$  \\
\hline
$10^{14}$ &  $2.3\,10^{-7}$ & $-1.4\,10^{-6}$  & $-7.3\,10^{-7}$ & $-2.9\,10^{-6}$  &  $9.4\,10^{-8}$ &  $-4.4\,10^{-7}$ &  $-1.6\,10^{-6}$  \\
\hline
$10^{15}$ &  $1.5\,10^{-7}$ & $-9.0\,10^{-7}$  & $-4.5\,10^{-7}$ & $-2.4\,10^{-6}$  &  $6.1\,10^{-8}$ &  $-5.7\,10^{-8}$ &  $-1.3\,10^{-6}$  \\
\hline
$10^{16}$ &  $1.6\,10^{-7}$ & $-9.3\,10^{-7}$  & $-4.0\,10^{-7}$ & $-2.6\,10^{-6}$  &  $6.1\,10^{-8}$ &  $3.7\,10^{-7}$ &  $-1.1\,10^{-6}$  \\
\hline
$10^{17}$ &  $1.3\,10^{-7}$ & $-7.3\,10^{-7}$  & $-2.6\,10^{-7}$ & $-3.1\,10^{-6}$  &  $4.9\,10^{-8}$ &  $8.6\,10^{-7}$ &  $-8.6\,10^{-7}$  \\
\hline
\end{tabular}\caption{\sl The values of the coefficients of the dimension six operators at $M_Z$. The values of the free parameters are the ones in Table~\ref{topdownvariation}. All $C_i$'s are in GeV${}^{-2}$.}\label{topdownCiexp}
\end{center}
\end{table}
One can notice that the values for the couplings in the two cases are not drastically different.

\subsection{Bottom-up running}

In this section we present the result of the running from low to high energy, with the parameters chosen to have the three coupling constants meet near a common value in the range $10^{14}-10^{17}$~GeV.  As in the previous subsection we first discuss the case in which the boundary condition for $\lambda$ is the one obtained from the running of the spectral action.

\subsubsection{Spectral action value for $\lambda$}

A good solution is one for which the common intersection is the starting point for the top-down running, and the $C_i$ come back to the original values given by the spectral action. We optimized our search for the unification, therefore the fact that the values of the $C_i$ ``come back'' to the same order within a factor of two or so, and are not off by an order a magnitude, is a check. The coefficient $C_{HWB}$ is not present at the $\Lambda$ scale in the spectral action, in this case one should expect it to be smaller than the other.
A further check is the value of the top Yukawa at $\Lambda$ which should be close to the value determined by the spectral action. The results for the coupling constants are in Table~\ref{unificationspectral}.  The quantities $\delta g_i(\%)$ indicate (in percent) how different is the value of the runned constants ($g_i^{run})$ with respect to the original spectral action value $g(\Lambda)$ we started with, as shown in Table~\ref{topdownvariation}.
\be
\delta g_i \%=\frac{|g_i^{run}(\Lambda-g(\Lambda)|}{g(\Lambda)} \times 100
\ee
with an analogous definition for $\delta y_t$.
\begin{table}[htp]
\begin{center}
\begin{tabular}{||c|c|c|c|c|c||}
\hline
$\Lambda$   &  $\delta g_1 \%$ &  $\delta g_2 \%$ &  $\delta g_3 \%$ &  $\delta y_t \%$  &  $\delta\lambda \%$    \\
\hline
$10^{14}$ & 1.4 & 2.1 & 0.17 & 0.30 &4.1 \\
$10^{15}$ & 3.3 & 0.02 & 0.54 & 3.6 & 4.7 \\
$10^{16}$ & 7.8 & 0.078 & 0.97 & 6.4 & 2.3\\
$10^{17}$ & 13 & 1.7 & 1.1 & 6.6 & 3.9 \\
\hline
\end{tabular}\caption{\sl The percent variation of the values of the three coupling constants and the top Yukawa coupling compared with the initial values of the top-down run.}\label{unificationspectral}
\end{center}
\end{table}

\begin{figure}[htb]
\begin{centering}
\includegraphics{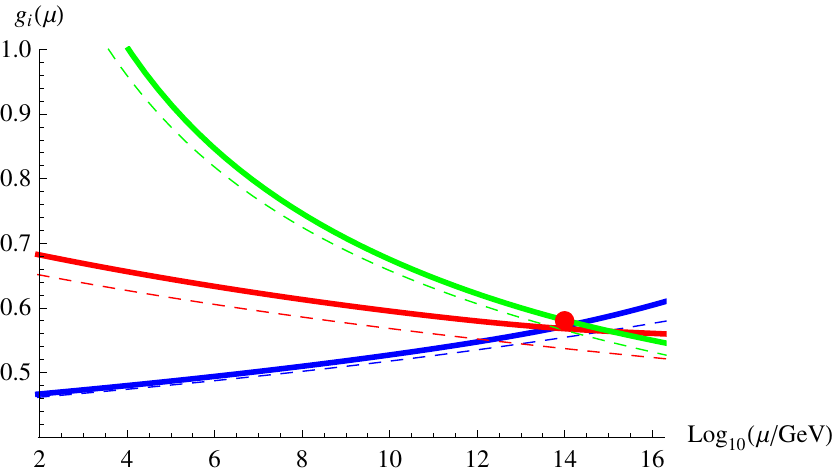}\includegraphics{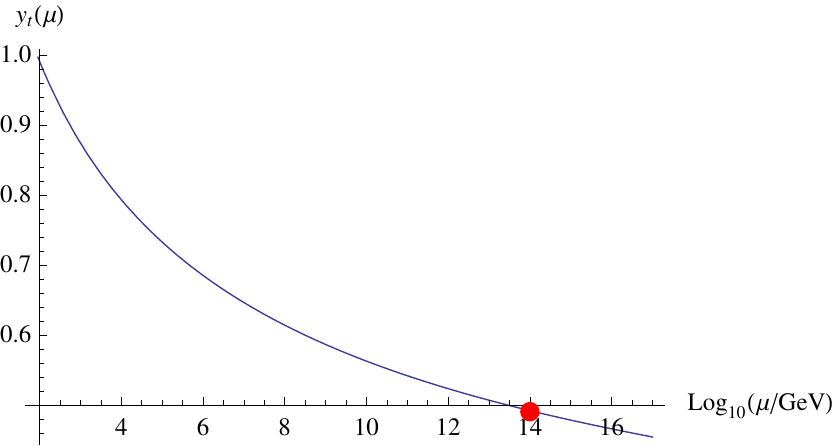}\end{centering}
\caption{\textsl{Running of the self-interaction
parameter $\lambda$(on the rigth) and gauge coupling constants (on
the left) in the presence of dimension six operators (thick lines)
and their standard behaviour (dashed lines) for $\Lambda=10^{14}$GeV. The values of the parameters are discussed in the text. The red dot indicates the starting value of the parameter. The dashed lines are the values of the $g_i$'s in the standard model.}} \label{figunif14}
\end{figure}

\begin{figure}[htb]
\begin{centering}
\includegraphics{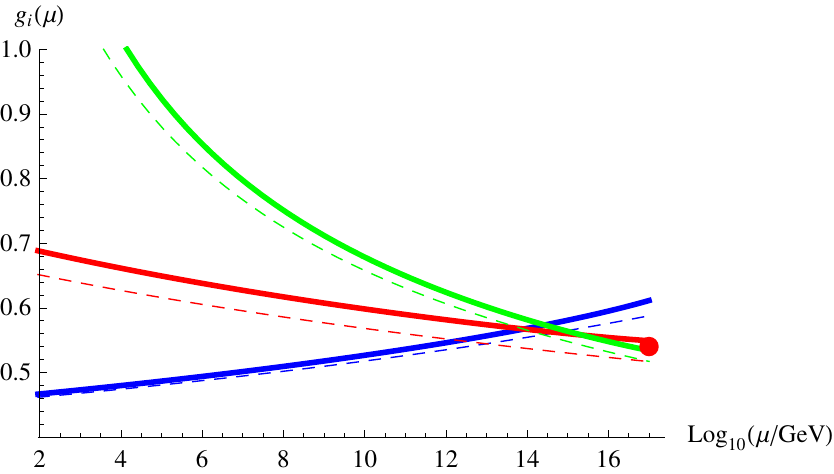}\includegraphics{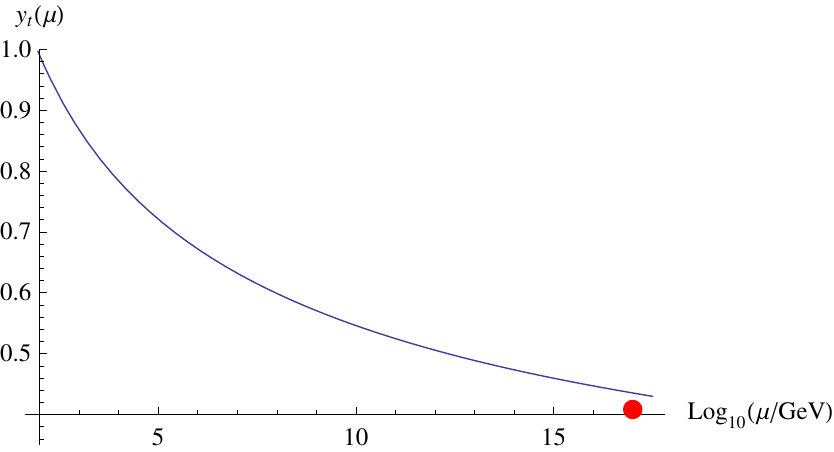}\end{centering}
\caption{\textsl{Same as in Fig.~\ref{figunif14} for $\Lambda=10^{17}$GeV.}} \label{figunif17}
\end{figure}

One can see that for the smaller values of $\Lambda\simeq 10^{14}-10^{15}$, one finds a good unification point, while for higher values the unification is worse. This can also be seen in Figs.~\ref{figunif14} and~\ref{figunif17} for the two extreme cases of $10^{14}$ and $10^{17}$ respectively, compared with the standard model running.
In the first case there is a good unification, while in the second case the point at which the constants meet is some way off the initial energy.
The values of the $C_i$'s at the scale $\Lambda$ are usually close to the one we started with in the top-down running, checking the consistency of the model. In particular $C_{HWB}$, which was zero, is constantly about one order of magnitude smaller than the other. We show this in Table~\ref{Ccomparison} for the two extreme values of $\Lambda$.
\begin{table}[htp]
\begin{center}
\begin{tabular}{||c|c|c|c|c|c|c|c||}
\hline
$\Lambda=10^{14}$  &  $C_{HWB}$ &  $C_{W}$ &  $C_{V}$ &  $C_{HV}$ &  $C_{H}$ &  $C_{HB}$ &  $C_{HW}$  \\
\hline
{\tiny Spec. Act.} &  $ 0 $ & $-3.0\,10^{-6}$  & $-3.0\,10^{-6}$ & $-5.2\,10^{-7}$  &  $-3.7\,10^{-6}$ &  $-8.1\,10^{-7}$ &  $-7.3\,10^{-7}$  \\
\hline
{\tiny Run} &  $1.3\,10^{-8}$ & $-1.5\,10^{-6}$  & $-1.6\,10^{-6}$ & $-5.5\,10^{-7}$  &  $-6.8\,10^{-6}$ &  $-6.7\,10^{-7}$ &  $-9.2\,10^{-7}$  \\ \hline  \hline
$\Lambda=10^{17}$  &  $C_{HWB}$ &  $C_{W}$ &  $C_{V}$ &  $C_{HV}$ &  $C_{H}$ &  $C_{HB}$ &  $C_{HW}$  \\
\hline
{\tiny Spec. Act.} &  $ 0 $ & $-4.2\,10^{-6}$  & $-4.2\,10^{-6}$ & $-5.4\,10^{-7}$  &  $6.9\,10^{-6}$ &  $-1.0\,10^{-6}$ &  $-9.4\,10^{-7}$  \\
\hline
{\tiny Run} &  $5.0\,10^{-8}$ & $-2.4\,10^{-6}$  & $-1.9\,10^{-6}$ & $-6.5\,10^{-7}$  &  $7.5\,10^{-6}$ &  $-8.7\,10^{-7}$ &  $-7.2\,10^{-7}$  \\
\hline
\end{tabular}\caption{\sl Comparison of the values of the coefficients of the dimension six operators at $\Lambda$. The second and fifth line are the initial values of the top-down running, as predicted by the spectral action for $\Lambda=10^{14}$. The third and the last lines refer to the $10^{17}$ case. All $C_i$'s are in GeV${}^{-2}$.}\label{Ccomparison}
\end{center}
\end{table}
Also in this case, the lower value for $\Lambda$ fares slightly better.

\subsubsection{Experimental value for $\lambda$}

\begin{figure}[htb]
\begin{centering}
\includegraphics{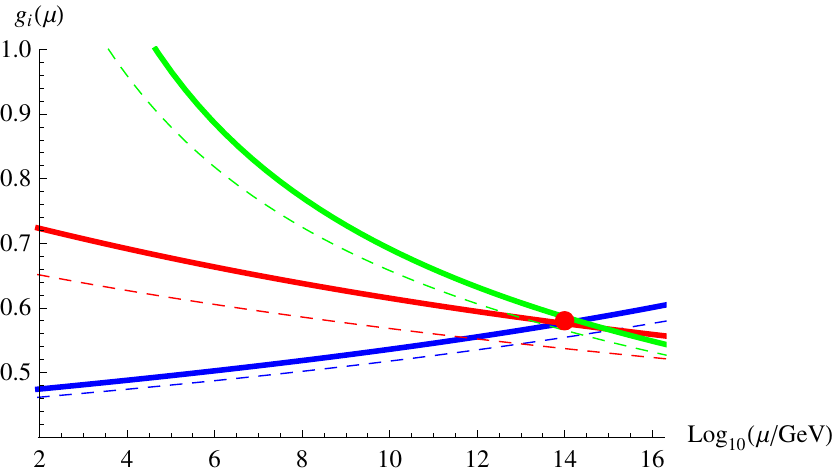}\includegraphics{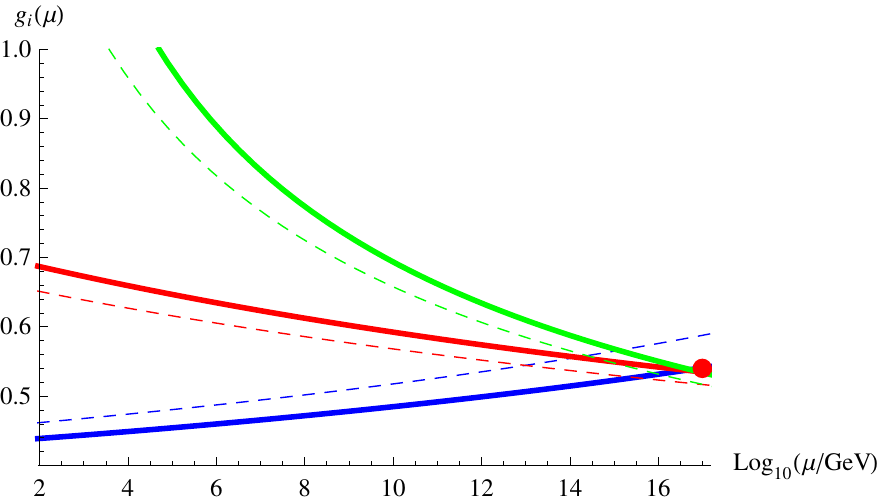}\end{centering}
\caption{\textsl{Gauge couplings unification  for two different unification scale $\Lambda=10^{14}$GeV (left) and $\Lambda=10^{17}$GeV (right) if one relaxes the spectral action boundaries.}} \label{figunif14and17exp}
\end{figure}

If one ignores the spectral action, and trusts it only in that it gives some boundary values for the dimension six operator coefficients, then the bottom-up running can be performed independently. In this subsection we present, therefore, the running of the coupling constants using as boundary conditions at $M_Z$ the experimental values for the $g_i$, $y_t$, $\lambda$, (eq.\ref{eq:InitialCondition},~\ref{expvalues}),  and the values of Table~\ref{topdownCiexp} for $C_i$'s and  we check if the unification is possible. As we can see from Fig.~\ref{figunif14and17exp},

for two different unification scales, the answer is positive if one relaxes the values of the dim-6 coefficients with respect to that suggested by the spectral action. In fact, in this case, the value of the  $\gamma$'s are slightly different from 1,  as shown  in table~\ref{topdownvariationexp}, but these allow to correct the unification point within an error of 1\%, as summarized in Table~\ref{unificationexp}.

\begin{table}[htp]
\begin{center}
\begin{tabular}{||c|c|c|c|c|c||}
\hline
$\Lambda(GeV)$   &  $\delta g_1 \%$ &  $\delta g_2 \%$ &  $\delta g_3 \%$  \\
\hline
$10^{14}$ & 0.62 & 0.74 &1.0  \\
$10^{15}$ & 1.4 & 0.38 & 0.56 \\
$10^{16}$ & 1.2 & 0.50 & 0.50 \\
$10^{17}$ & 0.14 & 0.98 & 1.1  \\
\hline
\end{tabular}\caption{\sl The percent variation of the values of the there coupling constants compared with the initial value of the unification point.}\label{unificationexp}
\end{center}
\end{table}

\section{Conclusions and Outlook}
In this paper we have calculated the sixth order terms appearing in the spectral action Lagrangian. We have then verified that the presence of these terms, with a proper choice of the free parameters, could cause the unification of the three constants at a high energy scale. Although the motivation for this investigation lies in the spectral noncommutative geometry approach to the standard model, the result can be read independently on it, showing that if the current Lagrangian describes an effective theory valid below the unification point, then the dimension six operator would play the proper role of facilitating the unification. In order for the new terms to have an effect it is however necessary to introduce a scale of the order of the TeV, which for the spectral action results in a very large second momentum of the cutoff function.

We note that we did not require a modification of the standard model spectral triple, although such a modification, and in particular the presence of the scale field $\sigma$, could actually improve the analysis.
From the spectral action point of view the next challenge is to include the ideas currently come form the extensions of the standard model currently being investigated. From the purely phenomenological side instead a further analysis of the effects of the dimension six operators for phenomenaology at large, using the parameters suggested by this paper, can be a useful pointer to new physics.

\noindent{\bf Acknowledgements} We thank Giancarlo D'Ambrosio and Giulia Ricciardi for discussions. DV was supported in parts by FAPESP, CNPq and by the INFN through
the Fondi FAI Guppo IV {\sl Mirella Russo} 2013. F.L. is partially supported by CUR Generalitat de Catalunya under projects FPA2013-46570 and 2014~SGR~104. A.D. and F.L. were partially supported by UniNA and Compagnia di San Paolo under the grant Programma STAR 2013. C.V.F. is supported by FAPESP.
\appendix
%dummy comment inserted by tex2lyx to ensure that this paragraph is not empty

\section{Spectral geometry \label{appspectralaction}}

\setcounter{equation}{0}

Noncommutative geometry is a way to describe noncommutative as well
as commutative space on equal footing. Being quite general, this approach
happens to be sufficiently rigid to make predictions about the standard
model. Noncommutative geometry uses many tools of spectral geometry.

Generally, the geometry of a noncommutative space is defined through a \textit{Spectral Triple} $\left(\mathcal{A},\mathcal{H},D\right)$
consisting of an algebra $\mathcal{A}$, a Hilbert space $\mathcal{H}$
and a Dirac operator $D$.

The algebra $\mathcal{A}$ should be thought of as a generalization
of the algebra of functions to the case when the underlying space is
possibly noncommutative.
The noncommutative algebra $\mathcal{A}$ relevant for the standard
model is the product of the ordinary algebra of functions on $\mathbb{R}^{4}$ times a finite dimensional matrix algebra $\mathcal{A}_{sm}=\mathbb{C}\oplus{\mathbb{H}}\oplus M_{3}(\mathbb{C})$. Here $\mathbb{H}$
is the algebra of quaternions, $M_{3}(\mathbb{C})$ is the algebra of
complex $3\times 3$ matrices. $\mathcal{A}_{sm}$ may be interpreted
as an algebra of functions on a finite "internal" space. Since just the internal part $\mathcal{A}_{sm}$ is noncommutative, the geometry corresponding
to Standard Model is called \textit{almost commutative}. The gauge group
is the group of automorphisms of $\mathcal{A}_{sm}$.

The algebra $\mathcal{A}$ acts on the Hilbert space $\mathcal{H}$.
We take $\mathcal{H}=L^{2}(\mathrm{sp}(\mathbb{R}^{4}))\otimes\mathcal{H}_{F}$ being a tensor product of the space of square-integrable spinors
and a finite-dimensional space $\mathcal{H}_F$.  The algebra $\mathcal{A}_{sm}$ acts on $\mathcal{H}_F$, and this imposes severe restrictions on
possible representations of the gauge group. Remarkably, these restrictions
are satisfied by the Standard Model fermions. To incorporate all of these
fermions one takes $\mathcal{H}_F=\mathcal{H}_R\oplus\mathcal{H}_L\oplus\mathcal{H}^c_R\oplus\mathcal{H}^c_L\oplus$, being $\mathcal{H}_R=\mathbb{C}^{24}$ ($8$ fermions with $3$ generations) the space of the right fermion, $\mathcal{H}_L=\mathbb{C}^{24}$ the space of the left fermions and the super index $c$ denotes their respectives antifermions. This give us the total of the $96$ SM degrees of freedom.

The Dirac operator $D$ also has to satisfy some consistency requirements,
that all are respected in the Standard Model. These conditions
\be
[\gamma,a]=0,\,\,\,\,[a,JbJ^{-1}]=0,\,\,\,\,[[D,a],JbJ^{-1}]=0
\ee
include the chirality operator $\gamma$ and the
real structure $J$.  Specifically, for the Standard Model the Dirac operator has the form
\begin{eqnarray}
D = D_M \otimes 1_{96} + \gamma^5 \otimes D_F,
\label{Diracoperator}
\end{eqnarray}
where $D_M=\gamma^\mu (\partial_\mu + \omega_\mu)$ is the canonical Dirac operator and the chirality and real structure are
$\gamma=\gamma^5\otimes\gamma_F$, $J=\mathcal{J}\otimes J_F$, with $\mathcal{J}$ being the charge conjugation. The finite dimensional Dirac operator $D_F$ is a matrix including the Yukawa couplings of leptons, Dirac and Majorana neutrinos. To introduce
he gauge fields and the Higgs we replace $D$ by
\be
D_A = D + \mathbb{A} + J\mathbb{A}J^{-1},
\ee
where $\mathbb{A}=\sum a[D,b]$, with $a,b\in \mathcal{A}$. In contrast to the usual Standard Model, the Noncommutative Standard Model
includes a singlet scalar field $\sigma$. Roughly speaking, this field is a result of "fluctuating" Majorana mass term of the Dirac operator and is responsible
for adjusting the Higgs' mass of noncommutative spectral action to the experimental values.
Some other approaches to the Higgs mass problem in spectral action can be found in
\cite{Stephan,coldplay,CCvS}.

Following Chamseddine and Connes tensorial notation\footnote{Not to be confused with the notation used in~\cite{coldplay}, where the meaning of dotted and undotted indices is different.} \cite{NCPart1}, an element $\Psi_M$ of the Hilbert space $\mathcal{H}$ is denoted as
\begin{eqnarray}
\Psi_M = \left(
\begin{array}{c}
\psi_{\alpha I}  \\
\psi_{\alpha' I'} \end{array}
 \right).
\label{notation}
\end{eqnarray}
First, the primed indices denote the conjugate spinor, this is $\psi_{\alpha' I'} = \psi^c_{\alpha I}$. The index $\alpha$ acts on $\mathbb{C}\oplus\mathbb{H}$ and is decomposed as $\alpha=\dot{a},a$ where $\dot{a}=\dot{1},\dot{2}$  acts on $\mathbb{C}$
and $a=1,2$ acts on $\mathbb{H}$. The index $I$ acts on $\mathbb{C}\oplus M_3(\mathbb{C})$ and is decomposed as $I=1,i\,$: the index $1$ acts on $\mathbb{C}$, that is another copy of the algebra of complex numbers,
and $i$ acts on $M_3(\mathbb{C})$.

Next, let us consider the action principle. One cannot construct too many
invariants by using the spectral triple data. One obvious choice is
the ordinary fermionic action
\begin{equation}
S_{F}=\langle \psi, D_{A}\psi\rangle \,.
\label{SF}
\end{equation}
As well, one can use the operator trace ${\rm Tr}$ in $\mathcal{H}$
to construct invariants from the Dirac operator alone.
In this way one obtains the \emph{Spectral action}
\begin{equation}
S_{\Lambda}\left(D\right)={\rm Tr}\, \left[ f \left(\frac{D^{2}}{\Lambda^{2}}\right)\right],\label{SLambda}
\end{equation}
where $f$ is a function restricted only by the requirement that trace
in (\ref{SLambda}) exists. $f$ is usually called the cutoff function
since it has to regularize (\ref{SLambda}) at large eigenvalues of $D$.
$\Lambda$ is a cutoff scale.

One can use the heat kernel expansion\footnote{See \cite{Gilkey02,Vassilevich} for a detailed overview of the heat trace
asymptotics.}
\begin{equation}
{\rm Tr}\, \left[ e^{-tD^2}\right] \simeq \sum_{p=0}^\infty
t^{-2+p} a_{2p}\bigl( D^2\bigr)\,,\qquad t\to+0,\label{heatex}
\end{equation}
to find a large $\Lambda$ expansion of the Spectral Action.
Suppose that $f$ is a Laplace transform,
\begin{equation}
f(z)=\int_0^\infty dt\,e^{-tz} \tilde f(t)\,.\label{Lap}
\end{equation}
Then
\begin{equation}
S_{\Lambda}\left(D\right)\sim\sum_{p=0}\Lambda^{4-2p}f_{2p}a_{2p}\left(D^{2}\right), \label{asymp}
\end{equation}
where
\begin{equation}
f_{2p}=\int_{0}^{\infty}dt\, t^{-2+p}\tilde f\left(t\right).
\end{equation}
Note, that we have restricted ourselves to four dimensions where the first four terms of the asymptotic
expansion are
\begin{equation}
S_{\Lambda}\left(D\right)\sim\Lambda^{4}f_{0}a_{0}\left(D^{2}\right)+\Lambda^{2}f_{2}a_{2}\left(D^{2}\right)+f_{4}a_{4}\left(D^{2}\right)+\frac{1}{\Lambda^{2}}f_{6}a_{6}\left(D^{2}\right)+\dots
\label{Sexp}
\end{equation}
$D^2$ is an operator of Laplace type. It can be represented as
\begin{equation}
D^2=-(\nabla^2+E)\,,\label{Laptype}
\end{equation}
where $\nabla$ is a covariant derivative, $\nabla_\mu=\partial_\mu +\omega_\mu$, $E$ is a zeroth order
term. Denoting by $\rm tr$ the usual matrix trace one may write
\begin{eqnarray}
a_{0} & = & \frac{1}{16\pi^{2}}\int d^{4}x\sqrt{g}{\rm tr}\, \left(1\right),\nonumber\\
a_{2} & = & \frac{1}{16\pi^{2}}\frac{1}{6}\int d^{4}x\sqrt{g}{\rm tr}\,\left[6E+R\cdot1\right],\nonumber\\
a_{4} & = & \frac{1}{16\pi^{2}}\frac{1}{360}\int d^{4}x\sqrt{g}
{\rm tr}\,
\left[\left(12R;_{\mu}^{\mu}+5R^{2}-2R_{\mu\nu}R^{\mu\nu}+2R_{\mu\nu\varrho\sigma}R^{\mu\nu\varrho\sigma}\right)\cdot1\right]\\
 & + & \frac{1}{16\pi^{2}}\frac{1}{360}\int d^{4}x\sqrt{g}{\rm tr}\,
 \left[60E;_{\mu}^{\mu}+60ER+180E^{2}+30\Omega_{\mu\nu}\Omega^{\mu\nu}\right].
\end{eqnarray}
Here $R_{\mu\nu\rho\sigma}$, $R_{\mu\nu}$ and $R$ are the Riemann tensor, the Ricci tensor and
the curvature scalar, respectively. The semicolon denotes covariant derivatives, and
$\Omega_{\mu\nu}=\partial_\mu \omega_\nu - \partial_\nu \omega_\mu +[\omega_\mu,\omega_\nu]$.

The expression for $a_{6}$ is rather long (see \cite{Vassilevich}), but it simplifies
if one considers a flat-space time
\begin{equation}
a_{6}^{flat}=\frac{1}{16\pi^{2}}\int d^{4}x\left(\Sigma_{\Omega}+\Sigma_{E}+\Sigma_{E\Omega}\right),\label{S01}
\end{equation}
where
\begin{eqnarray}
\Sigma_{\Omega} & = & {\rm tr}\,\left[-\frac{1}{90}\Omega_{\mu\nu;\tau}\Omega^{\mu\nu;\tau}+\frac{1}{180}\Omega_{\;\;\;\;;\nu}^{\mu\nu}\Omega_{\mu\rho}^{\;\;\;\;;\rho}
-\frac{1}{30}\Omega_{\mu\nu}\Omega^{\nu\tau}\Omega_{\tau}^{\;\;\mu}\right],\label{S02}\\
\Sigma_{E\Omega} & = & {\rm tr}\,\left[\frac{1}{12}E\Omega_{\mu\nu}\Omega^{\mu\nu}\right],\label{S03}\\
\Sigma_{E} & = & {\rm tr}\,\left[-\frac{1}{12}E^{;\mu}E_{;\mu}+\frac{1}{6}E^{3}\right],\label{S04}
\end{eqnarray}

Using the notation introduced in (\ref{notation}), the matrix elements of the connection $\omega_\mu$ are given by
\begin{eqnarray}
\left(\omega_{\mu}\right)_{\dot{1}1}^{\dot{1}1} & = & 0,\nonumber\\
\left(\omega_{\mu}\right)_{\dot{2}1}^{\dot{2}1} & = & ig_{1}B_{\mu}\otimes1_{4}\otimes1_{3},\nonumber\\
\left(\omega_{\mu}\right)_{b1}^{a1} & = & i \left(\frac{g_{1}}{2}B_{\mu}\delta_{b}^{a}-\frac{g_{2}}{2}W_{\mu}^{\tau}\left(\sigma^{\tau}\right)_{b}^{a}\right)\otimes1_{4}\otimes1_{3},\nonumber\\
\left(\omega_{\mu}\right)_{\dot{1}j}^{\dot{1}i} & = & i \left(-\frac{2g_{1}}{3}B_{\mu}\delta_{j}^{i}-\frac{g_{3}}{2}V_{\mu}^{m}\left(\lambda^{m}\right)_{j}^{i}\right)\otimes1_{4}\otimes1_{3},\nonumber\\
\left(\omega_{\mu}\right)_{\dot{2}j}^{\dot{2}i} & = & i \left(\frac{g_{1}}{3}B_{\mu}\delta_{j}^{i}-\frac{g_{3}}{2}V_{\mu}^{m}\left(\lambda^{m}\right)_{j}^{i}\right)\otimes1_{4}\otimes1_{3},\nonumber\\
\left(\omega_{\mu}\right)_{bj}^{ai} & = & i \left(-\frac{g_{1}}{6}B_{\mu}\delta_{b}^{a}\delta_{j}^{i}-\frac{g_{2}}{2}W_{\mu}^{\tau}\left(\sigma^{\tau}\right)_{b}^{a}
\delta_{j}^{i}-\frac{g_{3}}{2}V_{\mu}^{m}\left(\lambda^{m}\right)_{j}^{i}\delta_{b}^{a}\right)\otimes1_{4}\otimes1_{3}.
\end{eqnarray}
Therefore, the components of the curvature $\Omega_{\mu\nu}$ are
\begin{eqnarray}
\left(\Omega_{\mu\nu}\right)_{\dot{1}1}^{\dot{1}1} & = & 0,\nonumber\\
\left(\Omega_{\mu\nu}\right)_{\dot{2}1}^{\dot{2}1} & = & i g_{1}B_{\mu\nu}\otimes1_{4}\otimes1_{3},\nonumber\\
\left(\Omega_{\mu\nu}\right)_{b1}^{a1} & = & i \left(\frac{g_{1}}{2}B_{\mu\nu}\delta_{b}^{a}-\frac{g_{2}}{2}W_{\mu\nu}^{\tau}\left(\sigma^{\tau}\right)_{b}^{a}\right)\otimes1_{4}\otimes1_{3},\nonumber\\
\left(\Omega_{\mu\nu}\right)_{\dot{1}j}^{\dot{1}i} & = & i \left(-\frac{2g_{1}}{3}B_{\mu\nu}\delta_{j}^{i}-\frac{g_{3}}{2}V_{\mu\nu}^{m}\left(\lambda^{m}\right)_{j}^{i}\right)\otimes1_{4}\otimes1_{3},\nonumber\\
\left(\Omega_{\mu\nu}\right)_{\dot{2}j}^{\dot{2}i} & = & i \left(\frac{g_{1}}{3}B_{\mu\nu}\delta_{j}^{i}-\frac{g_{3}}{2}V_{\mu\nu}^{m}\left(\lambda^{m}\right)_{j}^{i}\right)\otimes1_{4}\otimes1_{3},\nonumber\\
\left(\Omega_{\mu\nu}\right)_{bj}^{ai} & = & i \left(-\frac{g_{1}}{6}B_{\mu\nu}\delta_{b}^{a}\delta_{j}^{i}-\frac{g_{2}}{2}W_{\mu\nu}^{\tau}\left(\sigma^{\tau}\right)_{b}^{a}\delta_{j}^{i}
-\frac{g_{3}}{2}V_{\mu\nu}^{m}\left(\lambda^{m}\right)_{j}^{i}\delta_{b}^{a}\right)\otimes1_{4}\otimes1_{3}.
\end{eqnarray}
The quantity $E$, which is defined through the operator $D^2$, has diagonal and non-diagonal components. The diagonal terms, i.e., $E_{\dot{1}1}^{\dot{1}1},E_{\dot{2}1}^{\dot{2}1},E_{b1}^{a1}$ and $E_{\dot{1}j}^{\dot{1}i},E_{\dot{2}j}^{\dot{2}i},E_{bj}^{ai}$
have the form
$E^{\rm diag}=-\frac{1}{2}\gamma^{\mu\nu}\Omega_{\mu\nu}-U\otimes1_{4}$, where
\begin{eqnarray}
U_{\dot{1}1}^{\dot{1}1} & = & \left(\left|y_{\nu}\right|^{2}\overline{H}H+\left|y_{\nu_{R}}\right|^{2}\sigma^{2}\right),\nonumber\\
U_{\dot{2}1}^{\dot{2}1} & = & \left(\left|y_{e}\right|^{2}\overline{H}H\right),\\
U_{b1}^{a1} & = & \left(\left|y_{e}\right|^{2}H_{a}\overline{H}^{b}+\left|y_{\nu}\right|^{2}\epsilon_{bc}\epsilon^{ad}\overline{H}^{c}H_{d}\right),\nonumber\\
U_{\dot{1}j}^{\dot{1}i} & = & \left(\left|y_{t}\right|^{2}\overline{H}H\right)\delta_{j}^{i},\nonumber\\
U_{\dot{2j}}^{\dot{2}i} & = & \left(\left|y_{d}\right|^{2}\overline{H}H\right)\delta_{j}^{i},\nonumber\\
U_{bj}^{ai} & = & \left(\left|y_{t}\right|^{2}H_{a}\overline{H}^{b}+\left|y_{d}\right|^{2}\epsilon_{bc}\epsilon^{ad}\overline{H}^{c}H_{d}\right)\delta_{j}^{i}.
\end{eqnarray}
and the non-diagonal components are
\begin{eqnarray}
E^{a1}_{\dot{1}1} & = & -\gamma^\mu \gamma_5 \otimes y^*_{\nu}\otimes \epsilon^{ab}\nabla_\mu H_b,\nonumber\\
E^{a1}_{\dot{2}1} & = & -\gamma^\mu \gamma_5 \otimes y^*_{e}\otimes \nabla_\mu \overline{H}^a,\nonumber\\
E^{\dot{1}1}_{a1} & = & -\gamma^\mu \gamma_5 \otimes y_{\nu}\otimes \epsilon_{ab}\nabla_\mu \overline{H}^b,\nonumber\\
E^{\dot{2}1}_{a1} & = & -\gamma^\mu \gamma_5 \otimes y_{e}\otimes \nabla_\mu H_a,\nonumber\\
E^{aj}_{\dot{1}i} & = & -\gamma^\mu \gamma_5 \otimes y^*_{t}\otimes \epsilon^{ab}\nabla_\mu H_b \delta^j_i,\nonumber\\
E^{aj}_{\dot{2}i} & = & -\gamma^\mu \gamma_5 \otimes y^*_{d}\otimes \nabla_\mu \overline{H}^a \delta^j_i,\nonumber\\
E^{\dot{1}j}_{ai} & = & -\gamma^\mu \gamma_5 \otimes y_{t}\otimes \epsilon_{ab}\nabla_\mu \overline{H}^b \delta^j_i,\nonumber\\
E^{\dot{2}j}_{ai} & = & -\gamma^\mu \gamma_5 \otimes y_{d}\otimes \nabla_\mu H_a \delta^j_i,
\end{eqnarray}
and the non-diagonal primed components
\begin{eqnarray}
E^{a'1'}_{\dot{1}1} & = & - y^*_{\nu_R} \overline{y^*_{\nu_R}}\otimes \epsilon^{ab} \overline{H}_b \sigma,\nonumber\\
E^{\dot{1}1}_{a'1'} & = & - y_{\nu_R} \overline{y^*_{\nu}}\otimes \epsilon_{ab} H^b \sigma,\nonumber\\
E^{\dot{1}'1'}_{a1} & = & - y^*_{\nu_R} y_{\nu} \otimes \epsilon_{ab} \overline{H}^b \sigma,\nonumber\\
E^{a1}_{\dot{1}'1'} & = & - y_{\nu_R} y^*_{\nu_R}\otimes \epsilon^{ab} H_b \sigma,\nonumber\\
E^{\dot{1}'1'}_{\dot{1}1} & = & -\gamma^\mu \gamma_5 \otimes y^*_{\nu_R}\otimes \partial_\mu \sigma,\nonumber\\
E^{\dot{1}1}_{\dot{1}'1'} & = & -\gamma^\mu \gamma_5 \otimes y_{\nu_R}\otimes \partial_\mu \sigma.
\end{eqnarray}

Taking just the contributions from $a_0$, $a_2$ and $a_4$ to the expansion (\ref{Sexp})
one reproduces quite well bosonic part of the Standard Model action, modulo the problem
with the Higgs mass and with the unification point that we have already mentioned above
\begin{eqnarray}
S &=& \int d^4x \left[- \frac{2}{\pi^2} f_2 \Lambda^2\left(\frac{1}{2}a\overline{H}H+\frac{1}{4}\left|y_{\nu_R}\right|^{2}\sigma^2\right)
+\frac{1}{2\pi^2}f_4 \left(\frac{5}{3}g^2_1B^2_{\mu\nu}+g^2_2W^2_{\mu\nu}+g^2_3V^2_{\mu\nu}+a(\nabla_\nu H)^2\right)\right]\nonumber\\
&+&\int d^4x \frac{1}{2\pi^2}f_4 \left[ b(\overline{H}H)^2 +2\left|y_{\nu}\right|^{2}\left|y_{\nu_R}\right|^{2}(\overline{H}H)\sigma^2+\frac{1}{2}\left|y_{\nu_R}\right|^{4}\sigma^4
+ \frac{1}{2}\left|y_{\nu_R}\right|^{2}(\partial_\mu \sigma)^2\right],\label{C01}
\end{eqnarray}
where
\begin{eqnarray}
a &=& \left|y_{\nu}\right|^{2}+\left|y_{e}\right|^{2}+3(\left|y_{t}\right|^{2}+\left|y_{d}\right|^{2}),\nonumber\\
b &=& \left|y_{\nu}\right|^{4}+\left|y_{e}\right|^{4}+3(\left|y_{t}\right|^{2}+\left|y_{d}\right|^{2})^2.
\end{eqnarray}

Higher order terms have been given considerably less attention. The papers \cite{WvS1,WvS2}
studied the influence of higher order terms on renormalizablity of Yang-Mills spectral
actions, while the works \cite{ILV1,ILV2,Kuliva} studied the spectral action beyond the
asymptotic expansion (\ref{asymp}).

The term $\Sigma_{\Omega}$ contains contributions from the gauge fields only
\begin{eqnarray}
\Sigma_{\Omega} & = & 8\left[\frac{1}{9}g_{1}^{2}B_{\mu\nu;\tau}^{2}+\frac{1}{15}g_{2}^{2}W_{\mu\nu;\tau}^{2}+\frac{1}{15}g_{3}^{2}V_{\mu\nu;\tau}^{2}\right]\nonumber \\
 & + & 8\left[-\frac{1}{18}g_{1}^{2}B_{\mu\nu;\nu}^{2}-\frac{1}{30}g_{2}^{2}W_{\mu\nu;\nu}^{2}-\frac{1}{30}g_{3}^{2}V_{\mu\nu;\nu}^{2}\right]\nonumber \\
 & + & 8\left[\frac{1}{10}g_{2}^{3}\varepsilon^{\delta\eta\kappa}W_{\mu\nu}^{\delta}W_{\nu\tau}^{\eta}W_{\tau\mu}^{\kappa}
 +\frac{1}{10}g_{3}^{3}f^{mnr}V_{\mu\nu}^{m}V_{\nu\tau}^{n}V_{\tau\mu}^{r}\right].\label{C02}
\end{eqnarray}
where:
\be
F_{\mu\nu;\tau}  =  \partial_{\tau}F_{\mu\nu}+ ig\left[ A_\tau, F_{\mu\nu}\right],
\ee
is the usual covariant derivative for gauge fields.

Note the presence of dimension six operators $X^{3}$ : $g_{2}^{3}\varepsilon^{\delta\eta\kappa}W_{\mu\nu}^{\delta}W_{\nu\tau}^{\eta}W_{\tau\mu}^{\kappa}$
and $g_{3}^{3}f^{mnr}V_{\mu\nu}^{m}V_{\nu\tau}^{n}V_{\tau\mu}^{r}$. Also note that the kinetic gauge terms, i.e, $F_{\mu\nu;\tau}^{2}$ and $F_{\mu\nu;\nu}^{2}$, are dimension six operators $X^2D^2$.
(We remind that $X$ is any field strength and by $D$ we indicate the fact that there are two derivatives).
Since under the trace and integral
\be
F_{\mu\nu;\tau}^{2}=2F_{\nu\mu;\nu}^{2}-4igF_{\mu\nu}F_{\nu\tau}F_{\tau\mu}\label{C02a},
\ee
these operators are not independent.

The term $\Sigma_{E\Omega}$ contains the operators $X^{2}H^{2}$:
\begin{eqnarray}
\Sigma_{E\Omega} & = & 8\left[\frac{g_{1}^{2}}{144}\left(15\left|y_{e}\right|^{2}+3\left|y_{\nu}\right|^{2}+17\left|y_{t}\right|^{2}+5\left|y_{d}\right|^{2}\right)
B_{\mu\nu}B_{\mu\nu}\left(\overline{H}H\right)\right]\nonumber \\
& + & 8\left[\frac{g_{2}^{2}}{48}\left(3\left|y_{t}\right|^{2}+3\left|y_{d}\right|^{2}+\left|y_{e}\right|^{2}+\left|y_{\nu}\right|^{2}\right)
W_{\mu\nu}^{\eta}W_{\mu\nu}^{\eta}\left(H\overline{H}\right)\right]\nonumber \\
& + & 8\left[\frac{g_{3}^{2}}{12}\left(\left|y_{t}\right|^{2}+\left|y_{d}\right|^{2}\right)V_{\mu\nu}^{m}V_{\mu\nu}^{m}\left(H\overline{H}\right)\right].\label{C03}
\end{eqnarray}

For the term $\Sigma_{E}$, let us write $\Sigma_{E}=\Sigma^0_{E}+\Sigma^{kin}_{E}+\Sigma^\sigma_{E}$, where
\begin{eqnarray}
\Sigma^0_{E}
&=&-8\left[\frac{1}{48}g_{1}^{2}\left(15\left|y_{e}\right|^{2}+3\left|y_{\nu}\right|^{2}+5\left|y_{d}\right|^{2}+17\left|y_{t}\right|^{2}\right)
H\overline{H}B_{\mu\nu}B^{\mu\nu}\right]\nonumber\\
&&-8\left[\frac{3}{48}g_{2}^{2}\left(\left|y_{e}\right|^{2}+\left|y_{\nu}\right|^{2}+3\left|y_{t}\right|^{2}+3\left|y_{d}\right|^{2}\right)
H\overline{H}W_{\mu\nu}^{\delta}W^{\delta\mu\nu}\right]\nonumber\\
&&-8\left[\frac{3}{12}g_{3}^{2}\left(\left|y_{t}\right|^{2}+\left|y_{d}\right|^{2}\right)H\overline{H}V_{\mu\nu}^{m}V^{m\mu\nu}\right]\nonumber \\
&&-8\left[\frac{1}{3}\left(\left|y_{e}\right|^{6}+\left|y_{\nu}\right|^{6}+3\left|y_{t}\right|^{6}+3\left|y_{d}\right|^{6}\right)
\left(\overline{H}H\right)^{3}\right] \nonumber\\
&&-8\left[\frac{1}{2}g_{2}^{3}\varepsilon^{\delta\eta\lambda}W_{\mu\nu}^{\delta}W_{\nu\tau}^{\eta}W_{\tau\mu}^{\lambda}
+\frac{1}{2}g_{3}^{3}f^{mnr}V_{\mu\nu}^{m}V_{\nu\tau}^{r}V_{\tau\mu}^{r}\right]\nonumber\\
&&-8\left[\frac{5}{12}g_{1}^{2}B_{\mu\nu;\tau}^{2}+\frac{3}{12}g_{2}^{2}W_{\mu\nu;\tau}^{2}+\frac{3}{12}g_{3}^{2}V_{\mu\nu;\tau}^{2}\right].
\label{C04}
\end{eqnarray}
is a contribution containing the operators: $X^2H^2$, $X^3$, $H^6$ and $X^2D^2$. The expression $\Sigma^{kin}_{E}$ contains higher order kinetic terms for the Higgs field, these are $H^4D^2$ and $H^2D^4$ operators:
\begin{eqnarray}
\Sigma^{kin}_E &=&-8\left[\frac{1}{2}\left(\left(\left|y_{\nu}\right|^{2}+\left|y_{e}\right|^{2}\right)^{2}+6\left|y_{t}\right|^{4}+6\left|y_{d}\right|^{4}\right)
\overline{H}H\left|\nabla_{\tau}H_{a}\right|^{2}\right]\nonumber\\
&&-8\left[\frac{1}{2}\left(\left(\left|y_{\nu}\right|^{2}-\left|y_{e}\right|^{2}\right)^{2}-3\left(\left|y_{t}\right|^{2}-\left|y_{d}\right|^{2}\right)^{2}\right)
\left|\overline{H}^{a}\nabla_{\tau}H_{a}\right|^{2}\right]\nonumber\\
&&-8\left[\frac{1}{12}\left(\left(\left|y_{e}\right|^{2}+\left|y_{\nu}\right|^{2}\right)^{2}+3\left(\left|y_{t}\right|^{2}+\left|y_{d}\right|^{2}\right)^{2}\right)
\left|\partial_{\tau}\left(\overline{H}H\right)\right|^{2}\right]\nonumber\\
&&-8\left[\frac{1}{6}\left(\left|y_{\nu}\right|^{2}+\left|y_{e}\right|^{2}+3\left|y_{t}\right|^{2}
+3\left|y_{d}\right|^{2}\right)\left(\partial_{\tau}\nabla_{\mu}\overline{H}^{a}\right)\left(\partial_{\tau}\nabla_{\nu}H_{a}\right)\right]\nonumber\\
&&-8\left[\frac{1}{12}\left(\left(\left|y_{e}\right|^{2}-\left|y_{\nu}\right|^{2}\right)^{2}
+3\left(\left|y_{t}\right|^{2}-\left|y_{d}\right|^{2}\right)^{2}\right)\left|\partial_{\tau}\left(H_{a}\overline{H}^{b}\right)\right|^{2}\right],
\label{C05}
\end{eqnarray}
and
\begin{eqnarray}
\Sigma^\sigma_E & = & -8\left[\frac{1}{2}\left|y_{\nu}\right|^{4}\left|y_{\nu_{R}}\right|^{2}\left(\overline{H}H\right)^{2}\sigma^{2}+
\frac{1}{2}\left|y_{\nu}\right|^{2}\left|y_{\nu_{R}}\right|^{4}\left(\overline{H}H\right)\sigma^{4}+\left|y_{\nu_{R}}\right|^{6}\sigma^{6}\right]\nonumber\\ &&-8\left[\frac{1}{6}\left|y_{\nu}\right|^{2}\left|y_{\nu_{R}}\right|^{2}\partial_{\tau}\left(\overline{H}^{a}\sigma\right)\partial_{\tau}
\left(H_{a}\sigma\right)+\frac{1}{12}\left|y_{\nu_{R}}\right|^{2}\left(\partial_{\mu}\partial_{\nu}\sigma\right)^{2}\right]\nonumber\\
&&-8\left[\frac{1}{2}\left|y_{\nu_{R}}\right|^{2}\left|y_{\nu}\right|^{2}\sigma^{2}\left|\nabla_{\tau}H_{a}\right|^{2}
+\frac{1}{2}\left|y_{\nu}\right|^{2}\left|y_{\nu_{R}}\right|^{2}\left(\sigma\partial^{\tau}\sigma\right)\nabla_{\tau}\left(\overline{H}H\right)\right]\nonumber\\ &&-8\left[\frac{1}{2}\left|y_{\nu}\right|^{2}\left|y_{\nu_{R}}\right|^{2}\left(\overline{H}H\right)\left(\partial_{\tau}\sigma\right)^{2}+
\frac{1}{2}\left|y_{\nu_{R}}\right|^{4}\left(\sigma\partial_{\tau}\sigma\right)^{2}\right]\nonumber\\
&&-8\left[\left|y_{\nu}\right|^{4}\left|y_{\nu_{R}}\right|^{2}\sigma^{2}\left(\overline{H}H\right)^{2}
+\frac{1}{2}\left|y_{\nu_{R}}\right|^{4} \left|y_{\nu}\right|^{2}\overline{H}H\sigma^{4}\right]\nonumber\\
&&-8\left[\frac{1}{12}\left|y_{\nu_{R}}\right|^{4}\left(\partial_{\tau}\sigma^{2}\right)^{2}+\frac{1}{6}\left|y_{\nu}\right|^{2}
\left|y_{\nu_{R}}\right|^{2}\partial_{\tau}\left(\overline{H}H\right)\left(\partial_{\tau}\sigma^{2}\right)\right],\label{C06}
\end{eqnarray}
is the contribution of the $\sigma$ singlet scalar field. As shown in (\ref{C01}), this field already appears in the $a_ 4$ coefficient. As has been explained in the main text, we do not consider this field
in the present work and, therefore, discard corresponding contributions to the action.

We consider that $y_{t}$, $y_\nu$ and $y_{\nu_R}$ are dominant and also define the variable $\rho$ as the ratio of the Dirac Yukawa couplings $y_{\nu}= \rho y_{t}$. Under this approximation and replacing (\ref{C02},\ref{C03},\ref{C04},\ref{C05}) in (\ref{S01}) we have
\begin{eqnarray}
a_{6}^{flat} &=& \frac{1}{2\pi^{2}}\int d^{4}x\left(\mathcal{O}_{X^2H^2}+\mathcal{O}_{H^6}+\mathcal{O}_{X^3}
+\mathcal{O}_{X^2D^2}+\mathcal{O}_{Kin}\right),\label{C07}
\end{eqnarray}
where
\begin{eqnarray}
\mathcal{O}_{X^{2}H^{2}} & = & -\frac{g_{1}^{2}}{72}\left(3\rho^{2}+17\right)\left|y_{t}\right|^{2}
B_{\mu\nu}B_{\mu\nu}\left(\overline{H}H\right)-\frac{g_{2}^{2}}{24}\left(3+\rho^{2}\right)\left|y_{t}\right|^{2}
W_{\mu\nu}^{\eta}W_{\mu\nu}^{\eta}\left(H\overline{H}\right)\nonumber \\
&&-\frac{g_{3}^{2}}{6}\left|y_{t}\right|^{2}V_{\mu\nu}^{m}V_{\mu\nu}^{m}\left(H\overline{H}\right).\label{C08}\\
\mathcal{O}_{H^6} & = & -\frac{1}{3}\left(\rho^{6}+3\right)\left|y_{t}\right|^{6}\left(\overline{H}H\right)^{3},\label{C09}\\
\mathcal{O}_{X^{3}} & = & -\frac{2}{5}g_{2}^{3}\varepsilon^{\delta\eta\lambda}W_{\mu\nu}^{\delta}W_{\nu\tau}^{\eta}W_{\tau\mu}^{\lambda}
-\frac{2}{5}g_{3}^{3}f^{mnr}V_{\mu\nu}^{m}V_{\nu\tau}^{r}V_{\tau\mu}^{r}\label{C10}.
\end{eqnarray}
are independent operators. $\mathcal{O}_{X^{2}D^{2}}$ is given by
\begin{eqnarray}
\mathcal{O}_{X^{2}D^{2}} & = & -\frac{11}{36}g_{1}^{2}B_{\mu\nu;\tau}^{2}-\frac{11}{60}g_{2}^{2}W_{\mu\nu;\tau}^{2}-\frac{11}{60}g_{3}^{2}V_{\mu\nu;\tau}^{2}\nonumber\\
&&+\frac{1}{18}g_{1}^{2}B_{\nu\mu;\nu}^{2}+\frac{1}{30}g_{2}^{2}W_{\nu\mu;\nu}^{2}+\frac{1}{30}g_{3}^{2}V_{\nu\mu;\nu}^{2},\label{C11}
\end{eqnarray}
and, as we have mentioned, they are dependent operators. The term $\mathcal{O}_{Kin}$ contains the $H^2D^4$, $H^4D^2$ operators.

As stated in \cite{Dim6Classification}, there are eight possible classes of dimension six operators: $XD^4$, $XH^2D^2$, $X^3$, $H^3$, $H^2X^2$, $H^4D^2$, $H^2D^4$ and $X^2D^2$. The first two classes $XD^4$, $XH^2D^2$ do not appear in the Spectral Action, since the unimodular condition excludes the terms proportional to $F_{\mu\nu}$. This condition also excludes any mixing between the gauge fields. The set of $X^3$, $H^3$, $H^2X^2$, $H^4D^2$ operators are independents. The class of $H^2D^4$ operators can be ``reduced'' to the independent $H^4D^2$. The operator $X^2D^2$ can be rewritten as a combination of $X^3$ and $H^4D^2$ operators. This can be done with the use of (\ref{C02a}): we write $F^2_{\mu\nu;\tau}$ in terms of $F_{\mu\nu}F_{\nu\tau}F_{\tau\mu}$ and $F^2_{\mu\nu;\mu}$, then we use the equation of motion for the gauge fields to obtain kinetic terms $H^4D^2$.

The operators $X^{3}$ and $X^{2}H^{2}$ affect the gauge coupling constants and the triple unification point. On the other hand, the operators $H^{6}$ and $\mathcal{O}_{kin}$ modify the Higgs mass. However, while $H^6$ produce a shift of order $C_H v^2 / \lambda$ (see eq. (\ref{Higgsmass})), the operators $\mathcal{O}_{kin}$  produce a shift of order $C_{kin}v^2$ (see \cite{JenkinsIII}) which will be negligible if $C_H$ and $C_{kin}$ are of the same order. Therefore, as a first approximation we will only focus on the RG contribution of the operators $X^{3}$, $X^{2}H^{2}$ and $H^6$
\begin{eqnarray}
a_{6}^{flat} = \frac{1}{2\pi^{2}}\int d^{4}x\left(\mathcal{O}_{X^2H^2}+\mathcal{O}_{H^6}+\mathcal{O}'_{X^3}\right),\label{C12}
\end{eqnarray}
where
\begin{eqnarray}
\mathcal{O}'_{X^{3}} =  -\frac{13}{60}g_{2}^{3}\varepsilon^{\delta\eta\lambda}W_{\mu\nu}^{\delta}W_{\nu\tau}^{\eta}W_{\tau\mu}^{\lambda}
-\frac{13}{60}g_{3}^{3}f^{mnr}V_{\mu\nu}^{m}V_{\nu\tau}^{r}V_{\tau\mu}^{r}\label{C13}.
\end{eqnarray}

Finally, in order to normalize the kinetic term $\frac{1}{2}a\left|\nabla_\mu H_a\right|^2$ in (\ref{C01}), we perform a rescaling
\be
H \rightarrow H= \sqrt{\frac{2}{3+\rho^2}} g \frac{H}{y_t},
\ee
where $g$ is the gauge coupling to the unification scale. In the end, we have
\begin{eqnarray}
\frac{f_6}{\Lambda^{2}}a_{6}^{flat} &=& \frac{1}{16\pi^{2}}\int d^{4}x\left(C_{HB}B^2_{\mu\nu}H\overline{H}+C_{HW}W^2_{\mu\nu}H\overline{H}+C_{HV}V^2_{\mu\nu}H\overline{H}\right)\nonumber\\
&+&\frac{1}{16\pi^{2}}\int d^{4}x\left(C_{H}(H\overline{H})^3+C_{W}\varepsilon^{\delta\eta\lambda}W_{\mu\nu}^{\delta}W_{\nu\tau}^{\eta}W_{\tau\mu}^{\lambda}
+C_{V}f^{mnr}V_{\mu\nu}^{m}V_{\nu\tau}^{r}V_{\tau\mu}^{r}\right)\label{eq:A6 Expression}
\end{eqnarray}
where
\begin{align}
C_{HB}(\Lambda) & =-\frac{f_{6}}{16\pi^{2}\Lambda^{2}}\frac{4\left(3\rho^2+17\right)}{9\left(\rho^2+3\right)}g^{4}\,\,,\, C_{HW}(\Lambda)=-\frac{f_{6}}{16\pi^{2}\Lambda^{2}}\frac{4}{3}g^{4}\,\,,\, C_{HV}(\Lambda)=-\frac{f_{6}}{16\pi^{2}\Lambda^{2}}\frac{16}{3\left(\rho^2+3\right)}g^{4}\,\,,\nonumber \\
C_{H}(\Lambda) & =-\frac{f_{6}}{16\pi^{2}\Lambda^{2}}\frac{512(\rho^{6}+3)}{3\left(\rho^2+3\right)^{3}}g^{6}\,\,,\,\,\,\,\, C_{W}(\Lambda)=-\frac{f_{6}}{16\pi^{2}\Lambda^{2}}\frac{26}{15}g^{3}\,\,,\,\,\,\,\, C_{V}(\Lambda)=-\frac{f_{6}}{16\pi^{2}\Lambda^{2}}\frac{26}{15}g^{3}
\end{align}

\end{document}